\documentclass[a4paper,11pt]{article}
\pdfoutput=1 

\usepackage{jcappub} 

\usepackage[T1]{fontenc} 

\newcommand{\lya}{Ly$\alpha$}

\newcommand{\mlya}{\mathrm{Ly}\alpha}
\newcommand{\IF}{I-front}
\newcommand{\odrt}{1D-RT}

\newcommand{\mIF}{\mathrm{IF}}
\newcommand{\HI}{H{\sc~i}}
\newcommand{\HII}{H{\sc~ii}}

\newcommand{\appropto}{\mathrel{\vcenter{
  \offinterlineskip\halign{\hfil$##$\cr
    \propto\cr\noalign{\kern2pt}\sim\cr\noalign{\kern-2pt}}}}}

\title{\boldmath Imaging reionization's last phases with I-front Lyman-$\alpha$ emissions}

\author[a]{Bayu Wilson,}
\author[a]{Anson D'Aloisio,}
\author[a]{George D. Becker,}
\author[b]{Christopher Cain,}
\author[c]{and Eli Visbal}

\affiliation[a]{Department of Physics and Astronomy, University of California, Riverside, CA 92521, USA.}
\affiliation[b]{School of Earth and Space exploration, Arizona State University, Tempe, AZ 85281, USA.}
\affiliation[c]{University of Toledo, Department of Physics and Astronomy and Ritter Astrophysical Research Center, Toledo, OH 43606.}

\emailAdd{bwils033@ucr.edu}
\emailAdd{ansond@ucr.edu}
\emailAdd{georgeb@ucr.edu}
\emailAdd{clcain3@asu.edu}
\emailAdd{elijah.visbal@utoledo.edu}

\abstract{
Long troughs observed in the $z > 5.5$ Ly$\alpha$ and Ly$\beta$ forests are thought to be caused by the last remaining neutral patches during the end phases of reionization -- termed neutral islands.  If this is true, then the longest troughs mark locations where we are most likely to observe the reionizing intergalactic medium (IGM).  A key feature of the neutral islands is that they are bounded by ionization fronts (I-fronts) which emit Lyman series lines. In this paper, we explore the possibility of directly imaging the outline of neutral islands with a narrowband survey targeting Ly$\alpha$. In a companion paper, we quantified the intensity of I-front Ly$\alpha$ emissions during reionization and its dependence on the spectrum of incident ionizing radiation and I-front speed. Here we apply those results to reionization simulations to model the emissions from neutral islands.  We find that neutral islands would appear as diffuse structures that are tens of comoving Mpc across, with surface brightnesses in the range $\approx 1 - 5\times 10^{-21}$ erg s$^{-1}$ cm$^{-2}$ arcsec$^{-2}$.  The islands are brighter if the spectrum of ionizing radiation driving the I-fronts is harder, and/or if the I-fronts are moving faster.  We develop mock observations for current and futuristic observatories and find that, while extremely challenging, detecting neutral islands is potentially within reach of an ambitious observing program with wide-field narrowband imaging. Our results demonstrate the potentially high impact of low-surface brightness observations for studying reionization.}

\begin{document}
\maketitle
\flushbottom

\section{Introduction}
\label{sec:intro}
Over the past ten years, quasar absorption studies have provided compelling evidence that reionization extended below $z=6$. The watershed toward this conclusion began with the confirmation of large fluctuations in the Ly$\alpha$ forest effective optical depth ($\tau_{\rm eff}$) at $z \gtrsim 5.5$ by Ref. \cite{2015MNRAS.447.3402B}, which expanded on the pioneering study of \cite{2006AJ....132..117F}. Especially striking was the discovery of an extreme $110$ $h^{-1}$cMpc trough in the Ly$\alpha$ forest of quasar ULAS J0148+0600 (\cite{2015MNRAS.447.3402B}; hereafter J0148).  Subsequent observations with larger quasar samples have significantly sharpened the measurements of the $\tau_{\rm eff}$ distribution \cite[e.g.][]{2018MNRAS.479.1055B, 2022MNRAS.514...55B, 2018ApJ...864...53E, 2020ApJ...904...26Y}.  It is now widely believed that the fluctuations are a signature of the tail of reionization, which is characterized by a mostly ionized IGM punctuated by the last remaining neutral regions in under-dense voids, termed neutral islands \cite{Kulkarni_2019, 2020MNRAS.491.1736K, Nasir_2020}. 

This model is supported by other observations as well. Ref. \cite{2018ApJ...860..155D} proposed to map the large-scale structure around the extreme J0148 trough, to determine whether its environment is over- or under-dense.  Observations have shown convincingly that the trough intersects a large-scale void in the galaxy distribution as traced by Ly$\alpha$ emitters (LAEs; \cite{2018ApJ...863...92B, 2021ApJ...923...87C, 2023ApJ...955..138C}) and by Lyman Break Galaxies (LBGs; \cite{2020ApJ...888....6K}). This finding is consistent with a model in which the long absorption troughs observed at $z\gtrsim 5.3$ are caused by neutral islands in the tail of reionization \cite{2020MNRAS.491.1736K, Nasir_2020}.  Another line of evidence is provided by measurements of the mean free path of ionizing photons. Refs. \cite{2021MNRAS.508.1853B,2023ApJ...955..115Z} observed a rapid evolution in the mean free path between $z = 5 - 6$, which can be recovered reasonably well in simulations where the global volume-averaged neutral fraction is still $\approx 20\%$ at $z=6$ \cite{cain_short_2021, 2021MNRAS.506.2390Q, 2022MNRAS.516.3389L, 2024MNRAS.530.5209R}. Reionization models with significant neutral fractions below $z=6$ are also consistent with updated constraints from the Ly$\alpha$ and Ly$\beta$ forest dark pixel fraction and the dark gap size distributions \cite{2021ApJ...923..223Z, 2022ApJ...932...76Z}. Lastly, there is strong observational evidence that dark gaps in the Ly$\alpha$ and Ly$\beta$ forests at $z < 6$ are associated with damping wing absorption, a hallmark of neutral islands \cite[e.g.][]{2015ApJ...799..179M}.  Damping wing absorption has been directly identified for the long trough towards J0148 \citep{2024arXiv240508885B}, as well as statistically for other long dark gaps \citep{2024arXiv240512275Z,2024arXiv240512273S}). 

Taken together, these results present a strong case for a reionization process extending below $z=6$. If correct, then the longest coeval troughs in the $z > 5$ Ly$\alpha$ and Ly$\beta$ forests mark locations where we are most likely to observe the reionizing IGM. In this paper, we explore one method to directly observe the reionization process that is hypothesized to be ongoing in sight lines such as J0148.  The neutral islands would be bounded by sharp I-fronts -- transition layers between the highly ionized and neutral IGM. Collisional excitations of H atoms are highly efficient within an I-front because of the close proximity of neutrals to hot electrons.  Hence I-fronts are sources of H line radiation, the brightest being the Ly$\alpha$ line.  In principle, the diffuse large-scale Ly$\alpha$ emissions tracing the I-fronts could provide a way to directly image the outline of a neutral island.   Although these emissions are expected to be faint, one feature motivating a dedicated effort to detect them is that we should know exactly where to aim the telescope(s) to find them. More broadly, the detection of even a single neutral island in this manner would be the first direct observation of a reionizing IGM -- a major empirical step forward toward understanding reionization and its sources. 

The current paper aims to develop mock images of neutral islands in Ly$\alpha$, spanning their expected range of intensities. This is itself a challenging task computationally. First, the problem requires an enormous dynamic range. One must simultaneously resolve the short spatial ($\sim 1$ pkpc) and temporal ($\Delta t < 1$ Myr) scales to model the I-front emissions accurately, as well as sample the large-scale structure of reionization for realistic neutral islands, which are expected to be $\sim 10~h^{-1}$cMpc across \cite{Nasir_2020}.  Second, one must include the effects of Ly$\alpha$ radiative transfer. Before reaching the observer, the emitted photons scatter off of neutral H atoms many times in the intervening IGM before redshifting out of resonance, which can change the morphology of the signal significantly.  Any realistic forecast must satisfy at least these requirements.  

 We take a multi-scale approach to tackling these difficulties.  This paper is the second in a two-part series.  In Ref.  \cite[hereafter Paper I]{2024_WilsonPaper1}, we used high-resolution one-dimensional radiative transfer (\odrt) simulations to quantify the parameter space of Ly$\alpha$ emissions over the range of plausible I-front characteristics during reionization. We found that the average Ly$\alpha$ production efficiency of an I-front, $\zeta$, defined to be the ratio of emitted Ly$\alpha$ flux to incident ionizing flux driving the front, depends mostly on just the spectrum of the ionizing radiation and the I-front speed ($v_{\rm IF}$). We characterize the former with a spectral index, $\alpha$. Paper I provides an interpolation table for $\zeta(v_{\rm IF}, \alpha)$, which serves as a kind of sub-grid model for applying our \odrt\ results to cosmological RT simulations of reionization with resolutions too coarse to model the I-front emissions directly.  In the current paper, we apply the sub-grid model from Paper I  to reionization simulations.  The simulations are then post-processed with a Ly$\alpha$ RT code to generate mock maps from a narrowband survey targeting Ly$\alpha$ at $z=5.7$ along a quasar sight line exhibiting long Ly$\alpha$/Ly$\beta$ forest troughs, the prototypical candidate being the sight line toward J0148.   We use our mocks to discuss prospects for detecting neutral islands with current and future observatories. 
 
 Refs. \cite{2023JCAP...05..041Y, 2023JCAP...05..042K} took a similar multi-scale approach to address a somewhat different problem. They applied results from a suite of high-resolution radiative transfer simulations to larger volume semi-numeric reionization simulations. They used these to explore the detectability of the power spectrum of polarized Ly$\alpha$ emission from I-fronts during reionization. Whereas they focused on the statistical detection of the emissions, we focus here on direct imaging of the neutral islands that are hypothesized to cause long troughs in the coeval Ly$\alpha$ and Ly$\beta$ forests at $z\sim 5.5 - 6$.  There are also number of ways that our methodology is complimentary to theirs, which are described in Paper I.

 The structure of this paper is as follows. In \S \ref{sect:methods} we describe our methodology to generate mock \lya\ emission maps from radiative transfer simulations of reionization. In \S \ref{sect:results} we present our main results.  We also develop mock observations to discuss current and future prospects for imaging neutral islands. We offer concluding remarks in \S \ref{sect:conclusion}. Throughout this work, we assume a standard $\Lambda$CDM cosmology with $\Omega_m$ = 0.3, $\Omega_\Lambda$ = $1-\Omega_m$, $\Omega_b$ = 0.048, and $h = 0.68$, consistent with the latest constraints \cite{2020A&A...641A...6P}. Distances are given in proper units unless otherwise noted.  For clarity, we will sometimes distinguish between proper and comoving units with a ``p'' and ``c'', respectively, e.g. pkpc or ckpc. 

\section{Methods} \label{sect:methods}
\subsection{Radiative transfer simulations of reionization}

Our models are based on simulations of reionization run with FlexRT, the ray tracing RT code of Ref. \cite{2024arXiv240904521C} (see also \cite{cain_short_2021, 2024MNRAS.531.1951C}). Ref. \cite{2024arXiv240904521C} contains a detailed code description as well as a battery of validation tests. We use FlexRT reionization simulations to construct mock sight lines containing neutral islands that are reasonably accurate in shape and size, and bounded by I-fronts moving at accurate speeds. The Ly$\alpha$ fluxes from the I-fronts are then assigned in post-processing using the efficiency tables of Paper I. 

FlexRT uses an adaptive ray tracing scheme on a uniform grid based on HealPix\footnote{\url{https://healpix.sourceforge.io}}, similar in spirit to the method described in Ref. \cite{2002MNRAS.330L..53A} and implemented in \cite{2007ApJ...671....1T}. The simulations were run in a cubic periodic volume with side $L=40~h^{-1}$cMpc and a uniform RT grid with $N=400^3$ cells, with cell size $\Delta x_{\rm cell} = 0.1~h^{-1}$cMpc.  These choices strike a balance between providing reasonable models of the large-scale shapes and sizes of neutral islands while at the same time allowing us to resolve small-scale features of the I-fronts bounding the islands. The RT simulations were run in post-processing on cosmological density fields extracted from an Eulerian hydrodynamics simulation run with the code of \cite{2004NewA....9..443T} at the same resolution as our RT grid. We saved density fields in snapshots spaced 10 Myr apart from $z=12-5.4$. Dark matter halos were identified with the same cadence using a spherical over-density criterion (see \cite{2015ApJ...813...54T} for more details).  Galaxies were populated using an abundance matching scheme described in \cite{2024MNRAS.531.1951C}. (Below, we will describe our model for the ionizing output of the galaxies.)  

Although FlexRT has multi-frequency RT capability (see Ref. \cite{2024MNRAS.531.1951C}), we opt here to run monochromatic simulations to reduce computational cost. Operating under the assumption that stellar emissions drove reionization, we adopt a source spectrum of $J_{\nu} \propto \nu^{-\alpha}$ between 1 and 4 Ry, and zero emissions at higher energies, to calculate an effective H ionization cross section 
 \begin{equation}\label{eq:sigma}
     \left<\sigma_\mathrm{HI}\right>(\alpha) = \int_{\nu_\mathrm{HI}}^{4\nu_\mathrm{HI}} d\nu \sigma_\mathrm{HI}(\nu)\phi(\nu,\alpha),
 \end{equation}
where the cross-section is well approximated as $\sigma_\mathrm{HI}(\nu) = \sigma_\mathrm{HI}^{\rm LL} \left(\frac{\nu}{\nu_{\rm LL}} \right)^{-2.75}$, $\sigma_\mathrm{HI}^{\rm LL}$ is the cross section at the Lyman limit, $\nu_{\rm LL}$, and $\phi \propto \nu^{-\alpha-1}$ is the normalized photon number spectrum, i.e. $\phi \propto \dot N$. We choose $\alpha = 1.5$, motivated by stellar population synthesis models of metal poor galaxies during reionization \cite{D_Aloisio_2019}. The effective monochromatic frequency, defined by $ \left<\sigma_\mathrm{HI}\right>(\alpha) = \sigma_\mathrm{HI}\left(\nu^\mathrm{eff}\right)$, is then $\nu^\mathrm{eff}= 19$ $\mathrm{ eV}/h_{\rm P}$ (where $h_{\rm P}$ is the Planck constant). The exact value of $\alpha$ used at this step for the reionization simulations is unimportant for our main results.  As highlighted in Paper I, the important quantity is the spectral index of ionizing radiation incident on the I-fronts bounding the neutral islands.  For this we will explore a range of values as described below.

One novel feature of FlexRT is its dynamical sub-grid model for the IGM opacity to ionizing photons, which has been used in Refs. \cite{cain_short_2021, 2024MNRAS.531.1951C}.   The model is built on an expanded suite of the small-scale radiative hydrodynamic simulations in \cite{2020ApJ...898..149D}.  However, their opacity model is not appropriate for the simulations in this paper because it was designed to work with $\Delta x = 1$ $h^{-1}$cMpc grid cells, much larger than our $\Delta x = 0.1$ $h^{-1}$cMpc RT cell size.  We instead assume that the cell-wise ionizing photon mean free path in ionized gas, $\lambda$ (the input of the sub-grid model of \cite{cain_short_2021}), is given by
\begin{equation}
    \lambda = \frac{1}{\sigma_{\rm HI} n_{\rm HI}} = \frac{\Gamma_{\rm HI}}{\sigma_{\rm HI}\alpha(T)(1+\chi)n_{\rm H}^2 },
\end{equation}
where $n_{\rm H}$ and $n_{\rm HI}$ are the total H and \HI\ number densities, respectively, $\Gamma_{\rm HI}$ is the \HI\ photo-ionization rate, $\alpha(T)$ is the recombination coefficient of \HII, and $\chi$ is the He to H number ratio.  The first expression is the definition of $\lambda$ in a uniform-$n_{\rm HI}$ gas, and the second equality comes from assuming highly ionized gas in photo-ionization equilibrium, which is usually appropriate behind I-fronts.  FlexRT uses the ``moving-screen'' I-front approximation, which assumes that each partially-ionized cell contains an infinitely sharp I-front.  In our simulations, I-fronts may span several cell widths, making this approximation (strictly speaking) incorrect.  However, our method of mapping Ly$\alpha$ fluxes integrated over I-fronts from the 1D RT calculations of Paper I (see \S \ref{sect:intrinsic_SB_IF}) would cause over-counting of Ly$\alpha$ flux if I-fronts were allowed to straddle multiple cells.  As such, the moving-screen approximation is most appropriate for our purposes.   The IGM temperature is tracked by applying the model of \cite{D_Aloisio_2019} for the temperature of the gas emerging from I-fronts, $T_{\rm reion}$. The subsequent evolution is modeled using eq. (6) of \cite{D_Aloisio_2019} as a sub-iteration of the RT time step. We track the most important heating and cooling mechanisms for the temperature of the highly ionized IGM: photoheating, cosmic expansion, Compton cooling, and collapse/expansion due to structure formation.\footnote{We also track recombination and free-free cooling, but these processes are subdominant at the temperatures and densities that are relevant here.}  We emphasize that the Ly$\alpha$ line emission from the I-fronts comes from the sub-grid model of Paper I, and does not depend on the temperatures tracked in the FlexRT reionization simulations.  These temperatures do, however, set the recombination radiation contribution to the Ly$\alpha$ flux. Both of these points will be elaborated below. 

\begin{figure}
    \centering
    \includegraphics[width=0.6\textwidth]{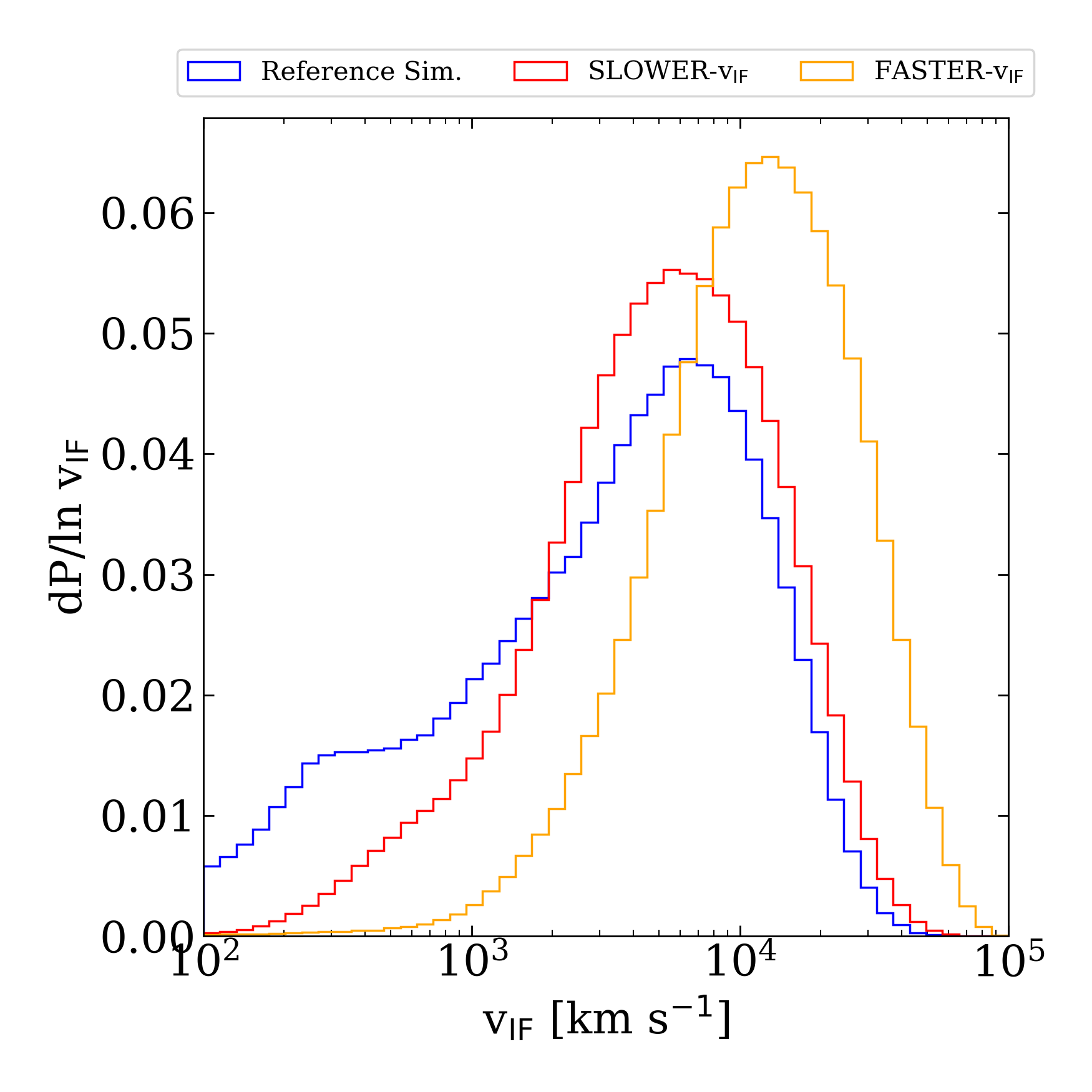}
    \caption{ \IF\ speed probability distributions from our reionization simulations (red and orange).  We compare against the distribution measured from the larger-volume simulation in Ref. \cite{2024MNRAS.531.1951C} at $z=5.7$ (blue).  We use this distribution as a reference for I-front speeds because the simulation has been calibrated to latch onto measurements of the mean Ly$\alpha$ flux at $z < 6$. The emissivity histories of our simulations are tuned to reproduce I-front speeds similar to the reference simulation. We employ two models, distinguished by their faster and slower I-front speeds, which will result in different I-front Ly$\alpha$ emissions.} 
    \label{fig:hist_vIF}
\end{figure}

In our reionization simulations we use a speed of light that is $30\%$ of the physical value, which reduces the required number of RT time steps by a factor of $3$, at the cost of unrealistically slow I-front speeds. Based on the results of Paper I, the I-front Ly$\alpha$ production efficiency is set entirely by the I-front speed ($v_{\rm IF}$) for fixed $\alpha$.  So achieving accurate $v_{\rm IF}$, in spite of our reduced speed of light approximation, is particularly important for our purposes. Our best empirical handle on I-front speeds near the end of reionization come from RT simulations calibrated to match the boundary conditions provided by measurements of the $z>5$ Ly$\alpha$ forest transmission, \cite[e.g.][]{2018ApJ...864...53E, 2020ApJ...904...26Y, 2022MNRAS.514...55B}. For one of the reionization simulations used in this paper, we tune the emissivity history of the source population such that the distribution of I-front speeds roughly matches the one measured in the ``reference'' simulation of Ref. \cite{2024MNRAS.531.1951C}.  This reference simulation had a larger volume with $L=200~h^{-1}$cMpc, used the full speed of light, and was carefully calibrated to latch onto the observed Ly$\alpha$ forest transmission.  So we believe that it provides a more realistic target model of I-front speeds near the end of reionization.  Given the large uncertainties, and to explore how the Ly$\alpha$ signal depends on I-front speed, we also construct a \textsc{faster-$v_\mathrm{IF}$} reionization model in which the I-fronts move a factor of 2 faster on average compared to the reference model. Motivating this choice, we found that, among the ($40~h^{-1}$cMpc)$^3$ sub-volumes of the reference simulation in \cite{2024MNRAS.531.1951C} that are $>10\%$ neutral, the mean I-front speeds can be up to a factor of 2 faster.

Figure \ref{fig:hist_vIF} compares the probability distribution functions (PDFs) of $v_{\rm IF}$ in our \textsc{slower-$v_\mathrm{IF}$} (red) and \textsc{faster-$v_\mathrm{IF}$} (orange) reionization models to the reference model (blue) from \cite{2024MNRAS.531.1951C} at $z=5.7$. Here and throughout this paper, we calculate the \IF\ speed in each partially-ionized RT cell (which by definition contains an unresolved I-front in our formalism) using 
\begin{equation}
    v_\mathrm{IF} = \frac{F^\mathrm{inc}_\mathrm{LyC}}{n_\mathrm{H}(1+\chi)},
    \label{eq:vIF}
\end{equation}
where $F^{\rm inc}_{\rm LyC}$ is the ionizing photon number flux incident on the I-front.\footnote{We have tested this against the reionization redshift gradient method used in  \cite{D_Aloisio_2019} and found good agreement.}  Note that the \textsc{slower-$v_\mathrm{IF}$} model has been calibrated such that the peak of its $v_{\rm IF}$ distribution approximately matches that of the reference distribution. This model may underestimate $v_{\rm IF}$ because neutral islands are expected to be found in under-dense voids and at a late stage of reionization in which there are many overlapping streams of ionizing radiation. The main takeaway from Figure \ref{fig:hist_vIF} is that the I-fronts bounding the neutral islands in our simulations have $v_{\rm IF}$ within the range of expectation, ultimately informed by empirical measurements of the $z>5$ Ly$\alpha$ forest transmission (by way of the calibrated reference simulation in Ref. \cite{2024MNRAS.531.1951C}).

\begin{figure}
    \centering
    \includegraphics[width=0.5\textwidth]{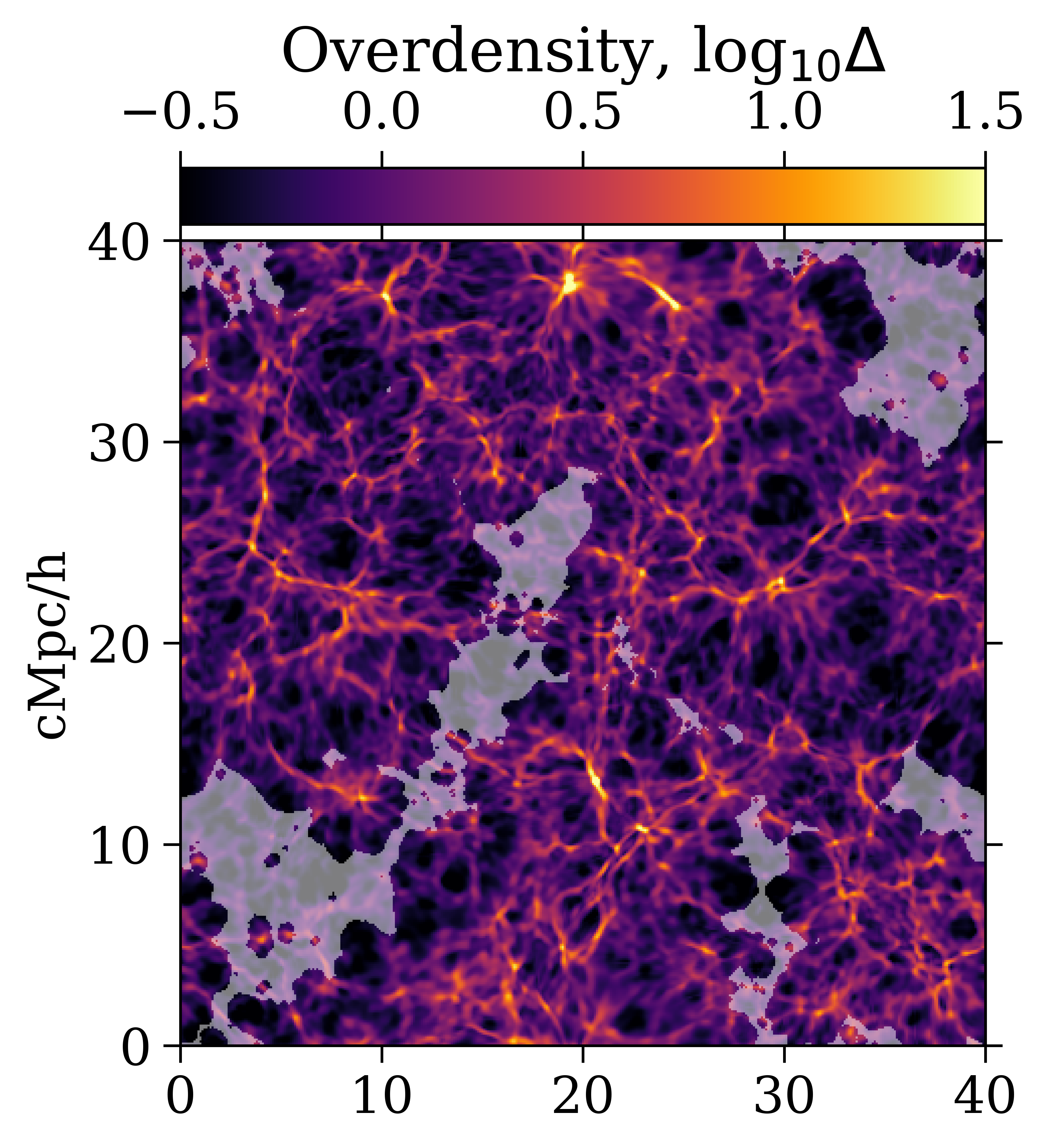}
    \caption{A slice through the gas density field from one of our reionization simulations at a global neutral fraction of $x_\mathrm{HI}=20\%$. The gas over-density, $\Delta$, is in units of the cosmic mean. The lightly shaded regions correspond to cells with neutral fraction $\geq 10\%$, which visualize the neutral islands (the locations of which are not sensitive to the exact choice of threshold). This paper quantifies the possibility of detecting Ly$\alpha$ emissions from I-fronts bounding neutral islands, such as the shaded regions in this plot.}
    \label{fig:gas_slices}
\end{figure}

Lastly, Figure \ref{fig:gas_slices} shows an example slice through the gas density field for one of our simulations at an epoch with volume-weighted average neutral fraction $x_\mathrm{HI}=20\%$. The lightly shaded regions correspond to cells with neutral fraction $\geq 10\%$, visualizing the neutral islands (the locations of which are not sensitive to the exact choice of threshold). The neutral islands are located in the voids of the density field near the completion of the reionization process.

\subsection{Ly$\alpha$ emissions and radiative transfer} \label{sect:intrinsic_SB_IF}

We apply the I-front Ly$\alpha$ emission results from Paper I to our reionization simulations in post processing. In Paper I, we showed that the Ly$\alpha$ production efficiency of an I-front depends on the spectral index of the ionizing flux driving the front. The spectrum of the ionizing radiation background during reionization is highly uncertain. For instance, Ref. \cite{2020A&A...634A.134G} report that stars stripped by binary interactions could contribute tens of percent of the ionizing photon budget, substantially hardening the spectrum of the ionizing background. Ref. \cite{D_Aloisio_2019} found that the spectral index of the background depends on the metallicities of the stellar populations. Making matters more complicated, the spectral index of the incident flux will vary from I-front to I-front because of spatial variations in the source properties, as well as spectral filtering from absorption by H atoms between the sources and I-front. Because of the $\sigma_{\rm HI} \sim \nu^{-2.75}$ frequency-dependence of the \HI\ photoionization cross section, most of the absorptions occur at energies near the Lyman limit, which acts to harden the spectrum. We do not attempt to model these effects here.  In appendix B of paper I, we used high-resolution \odrt\ simulations to explore the expected degree of spectral hardening. We found a characteristic hardening of $\Delta \alpha \sim -0.75$ across a path length of $\sim 15~h^{-1}$cMpc in the IGM at $z\sim 5.7$.\footnote{We note, however, that the spectral hardening is sensitive to the highly uncertain column density distribution at these redshifts.} In Appendix \ref{appendix:hardening} of the current paper, we provide a complimentary calculation based on a multi-frequency FlexRT simulation in a $L=40~h^{-1}$cMpc box, which corroborates this degree of spectral hardening.

\begin{figure*}
    \centering
    \includegraphics[width=1.02\textwidth]{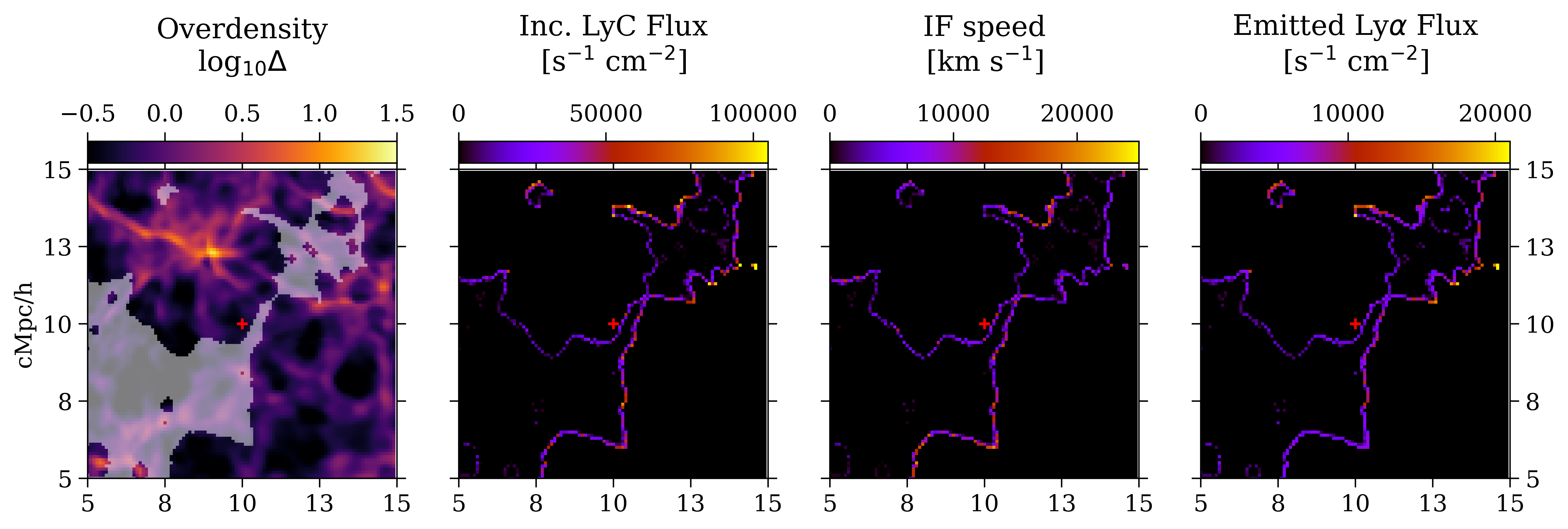}
    \caption{A zoom-in on a neutral island located near $(x,y) \approx (10, 10)~h^{-1}$cMpc in Figure \ref{fig:gas_slices}. The left panel shows the cosmological density field, with neutral regions shaded in gray, while the center-two panels show \IF\ properties relevant for \lya\ emission: the ionizing flux incident on the I-front and the I-front speed. The right panel shows the emitted \lya\ flux, obtained from the model in Paper I, which is based on a suite of high-resolution \odrt\ simulations. We assume an incident spectral index of $\alpha=0.5$. These panels illustrate our process for modeling the emitted Ly$\alpha$ flux from I-fronts. The emitted flux is then processed with Ly$\alpha$ RT to get the observed flux (not shown here -- see Figure \ref{fig:SB_and_LyaRT}). Note that our full calculations also include Ly$\alpha$ recombination radiation from the cosmic web, which is not shown here.} 
    \label{fig:zoom_slice}
\end{figure*} 

In what follows, we choose two representative values for the spectral index: $\alpha = 1.5$ and $\alpha = 0.5$. The former is motivated by population synthesis models of metal-poor stellar populations during reionization \cite{D_Aloisio_2019}. The latter represents a scenario in which the spectrum of ionizing radiation driving the I-fronts is harder owing to intrinsically harder source spectra \cite[e.g.][]{2020A&A...634A.134G} and/or to IGM filtering. Based on the results of Paper I, the $\alpha = 0.5$ model will generally yield brighter I-front Ly$\alpha$ emissions.  We calculate I-front speeds using eq. (\ref{eq:vIF}). For fixed $\alpha$, $v_{\rm IF}$ can be mapped directly to a Ly$\alpha$ production efficiency, $\zeta(v_{\rm IF}, \alpha)$, from the interpolation table made publicly available in Paper I.  The efficiency is defined to be 

\begin{equation}
\zeta(v_{\rm IF}, \alpha) = \frac{F^{\rm emit}_{\mathrm{Ly}\alpha}}{F^{\rm inc}_{\rm LyC}},
\label{eq:efficiency}
\end{equation}
where $F^{\rm emit}_{\mathrm{Ly}\alpha}$ is the number flux of Ly$\alpha$ photons emitted from the I-front. Multiplying eq. (\ref{eq:efficiency}) by $F^{\rm inc}_{\rm LyC}$ gives the Ly$\alpha$ flux emitted from the RT cell containing the I-front. 

In Figure \ref{fig:zoom_slice}, we illustrate this procedure by showing a zoom-in of the neutral island located near $(x,y) \approx (10,10)~h^{-1}$cMpc in Figure \ref{fig:gas_slices}. The slices are all one cell thick ($\Delta x_{\rm cell} = 0.1~h^{-1}$cMpc). The center two panels show $F^\mathrm{inc}_{\rm LyC}$ and $v_\mIF$ along the boundary of the neutral island, while the right panel shows the resulting \lya\ flux, $F^{\rm emit}_{\mathrm{Ly}\alpha}$, assuming an incident spectral index of $\alpha=0.5$. Variations in $F^\mathrm{inc}_{\rm LyC}$ owe to spatial variations in the ionizing radiation background, while variations in $v_{\rm IF}$ may owe to either those in $F^\mathrm{inc}_{\rm LyC}$ or in the local density. 

Although not the focus of this paper, we also include Ly$\alpha$ emissions from recombinations in our main results below, to show that these do not wash out the I-front signal. For the IGM at high-$z$, we adopt the usual approximation, valid in the limit where the opacity to Lyman series radiation is large, that every emission from radiative recombinations to energy levels with $n \geq 3$ will quickly be processed down to either a Ly$\alpha$ line photon ($2P\rightarrow 1S$) or two-photon emission ($2S\rightarrow 1S$). We use the case B recombination coefficient fit from Ref. \cite{Hui_1997} and the fitting function from \cite{Cantalupo_2008} for the fraction of recombinations to $n > 1$ yielding a Ly$\alpha$ photon.  (This fraction is only mildly dependent on temperature and is typically around $68\%$.) 

The above procedure yields the {\it intrinsic} Ly$\alpha$ flux emitted by a given cell -- from both I-fronts and radiative recombinations.  We model the scattering of this radiation by neutral hydrogen in the IGM using the 3D Monte Carlo (MC) \lya\ RT code of Ref. \cite{Visbal_2018}, which is based on the method detailed in the appendix of \cite{2010ApJ...725..633F}.  Each RT cell is a source from which we inject MC photon packets with probability proportional to the cell \lya\ emissivity. The injected packets are given random directions and their starting frequencies are drawn from a Gaussian line function with Doppler parameter set by the gas temperature of the cell. For each photon packet, a reference optical depth, $\tau$, is drawn randomly from the distribution $P(\tau) \propto \exp(-\tau)$. This reference $\tau$ marks the next scattering event for the packet.  The optical depth to Ly$\alpha$ scattering, eq. (C2) in \cite{2010ApJ...725..633F}, is tracked for the packet until it reaches the reference $\tau$, at which point the packet is scattered. The new frequency of the packet is calculated using eq. (C3) in \cite{2010ApJ...725..633F} and the new direction is picked from a dipolar phase function which is also in Appendix C of \cite{2010ApJ...725..633F}.  To save computational time, we pre-compute the integral appearing in eq. (C2) of Ref. \cite{2010ApJ...725..633F} at 200 temperatures logarithmically spaced from $10^2$ K to $1.3\times10^5$ K and $10,000$ dimensionless frequencies, $x=\frac{\nu-\nu_0}{\Delta \nu_D}$, ranging from 0 to 10, where $\Delta \nu_D$ is the Doppler width. The repeated scattering of the MC photons results effectively in their random walk through position and frequency space.  The Hubble flow causes the \lya\ photons to redshift out of resonance after traveling a few to tens of cMpc/h. We tile together copies of our simulation volume (with $L = 40~h^{-1}$cMpc) using periodic boundary conditions to track scatterings out to a distance $160~h^{-1}$cMpc.  By the time the photons have traveled this pathlength we can safely assume that they have redshifted out of resonance. For each run, we inject $6\times10^5$ MC photons, which we find sufficiently resolves the observed I-front emissions for our purposes.

\section{Results}\label{sect:results}
\subsection{Surface Brightness maps}

\begin{figure*}
    \centering
    \includegraphics[width=1.0\textwidth]{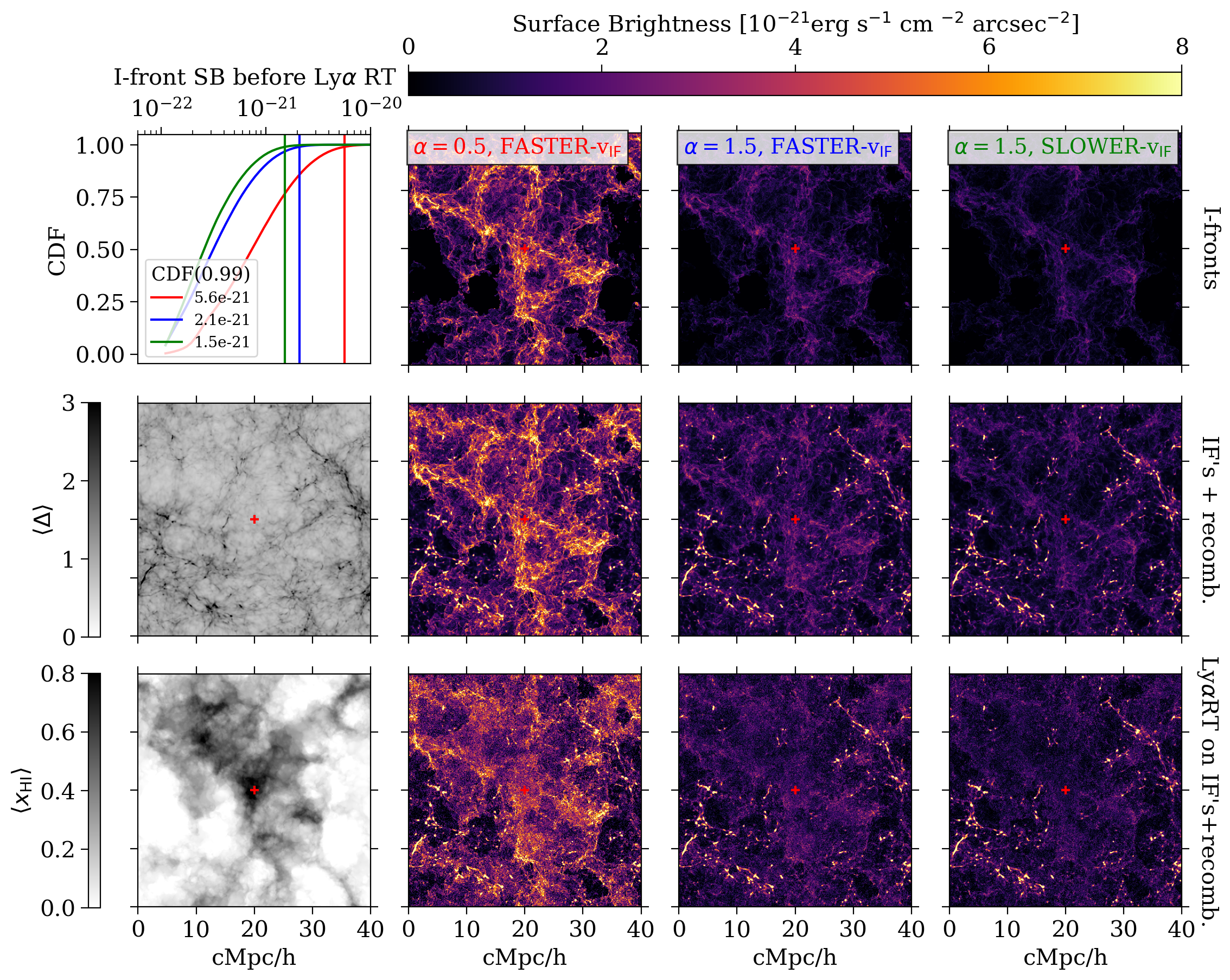}
    \caption{Simulated narrowband \lya\ surface brightness maps for three reionization models at $z=5.7$. The maps are centered on a highly neutral region, qualitatively similar to the neutral islands that are hypothesized to generate long troughs observed in the $z>5$ Ly$\alpha$ and Ly$\beta$ forests.  This can be seen in the bottom-left panel, which shows the neutral fraction averaged over the narrowband filter depth (see main text). For reference, the middle-left panel shows the gas density in units of the cosmic mean.  The top row shows the I-front emissions only.  The middle row adds in the recombination radiation.  The bottom row includes the effects of Ly$\alpha$ RT.  As indicated in the map legends, the models vary the spectral index of the ionizing radiation driving the I-fronts, as well as the I-front speeds. To aid comparison, the top-left panel shows the surface brightness cumulative distribution functions (CDFs) of the I-front surface brightness, not including the effects of \lya\ RT. The colors of the lines match the colors given in the map legends. The vertical lines mark the 99th percentiles for each model. In this paper we consider the possibility of imaging neutral islands directly using the Ly$\alpha$ radiation from the I-fronts bounding them.}
    \label{fig:SB_and_LyaRT}
\end{figure*}

The mode of observation that we focus on in this paper is narrowband imaging by a ground based observatory around quasar sight lines suspected of containing neutral islands (owing to the presence of long Ly$\alpha$/Ly$\alpha$ beta forest troughs). The observed surface brightness may be estimated from our simulated intensity by integrating in frequency/redshift over the width of the narrowband filter. We choose an integration distance (along the line of sight) that is motivated by the NB816 filter of the Hyper-Suprime Cam (HSC) on the Subaru Telescope, which has been used to identify Ly$\alpha$ emitting galaxies at $z=5.7$ \cite[e.g.][]{2018PASJ...70S..13O, 2021ApJ...923...87C, 2023ApJ...955..138C}. The filter has $>50\%$ peak transmission over observer-frame $8122 < \lambda < 8239$ \AA, which we choose as our integration interval.  For reference, this corresponds to $\Delta r_\mathrm{comoving} \approx 26$ $h^{-1}$cMpc, or $\approx 65\%$ of our simulation box. 

Figure \ref{fig:SB_and_LyaRT} shows $z=5.7$ \lya\ surface brightness maps from our simulations.  We center the maps on a neutral island seen in the bottom-left panel, which shows the average neutral fraction projected over the approximate filter width of $26~h^{-1}$cMpc. (The red cross marking the center of the map can be seen in all panels.)  The neutral gas is mainly relegated to the under-dense voids as indicated in the panel above, which shows the gas density projected over the same distance. The surface brightness maps in the top row show the emitted Ly$\alpha$ brightness, from I-fronts only, for three models,  i.e. these maps do not include the effects of Ly$\alpha$ radiative transfer, nor do they contain recombination radiation.  The model labeled \textsc{$\alpha = 0.5$, faster-$v_\mathrm{IF}$} (2nd column) assumes a harder spectral index and a reionization history yielding the faster I-front speeds shown in Figure \ref{fig:hist_vIF}.  This model is our most optimistic in the sense that the harder spectrum and faster I-front speeds generally lead to brighter I-front emissions.  The model labeled 
\textsc{$\alpha = 1.5$, faster-$v_\mathrm{IF}$} (3rd column) has the same I-front speeds, but assumes a softer spectrum with $\alpha = 1.5$.  Lastly, the \textsc{$\alpha = 1.5$, slower-$v_\mathrm{IF}$} model also assumes $\alpha = 1.5$, but uses the reionization model with the slower I-front speeds in Figure \ref{fig:hist_vIF}.  This model exhibits the dimmest I-front emissions.  To aid the comparison between models, the top-left panel provides the cumulative distribution function (CDF) of I-front surface brightnesses for the three models, with color scheme matching the labels in the top row.  The vertical lines mark the 99th percentiles of each model.  Note that the CDF does not include the effects of Ly$\alpha$ RT.

Comparing the $\alpha=0.5$ and $\alpha=1.5$ models for fixed I-front speeds, and referring to the provided CDF, we find that the I-fronts are roughly a factor of 2.6 brighter in the former.  This owes to the wider I-fronts and hotter internal temperatures generated by harder ionizing spectra, which together yield more total I-front emission (see Paper I for a more detailed discussion).  Similarly, comparing the models with fixed $\alpha = 1.5$ but with faster and slower I-front speeds, we find that the I-fronts are a factor of $\approx 1.4$ brighter in the former. The emitted Ly$\alpha$ flux is proportional to the incident ionizing flux, which is proportional to the I-front speed (c.f. eqs. \ref{eq:efficiency} and \ref{eq:vIF}).  Because of this proportionality, faster I-front speeds generally yield brighter I-front emissions. This is true despite the fact that the efficiency, $\zeta$, decreases with $v_{\rm IF}$. In other words, the larger $F^{\rm inc}_{\rm LyC}$ ``wins'' over the suppressed $\zeta$. Note that we have omitted an \textsc{$\alpha=0.5$, slower-$v_{\rm IF}$} scenario for brevity. Qualitatively, we find that this scenario falls between the \textsc{$\alpha = 0.5$-faster-$v_{\rm IF}$} and \textsc{$\alpha = 1.5$-faster-$v_{\rm IF}$} models.\footnote{This can be deduced by examining the PDFs of $v_{\rm IF}$ in Figure \ref{fig:hist_vIF}, and using the results from Paper I to estimate the boost in surface brightness in going from $\alpha = 1.5$ to $\alpha = 0.5$.}

The middle row of Figure \ref{fig:SB_and_LyaRT} adds in the Ly$\alpha$ radiation produced by recombinations to the intrinsic emission maps (without the effects of Ly$\alpha$ RT).  The recombination radiation reaches a similar peak brightness to the I-front emissions in the \textsc{$\alpha = 0.5$, faster-$v_\mathrm{IF}$} model, but only in high density gas near the peaks of the density field. This contrasts with the I-front emissions, which are largely coming from the under-dense regions of the IGM, because that is where the neutral islands reside near the end of reionization. (Compare to the density field shown in the middle-left panel.) The recombination radiation is comparable to or
brighter than the I-front emissions in our other two models.\footnote{For reference, from left to right in the middle row of Figure \ref{fig:SB_and_LyaRT}, the total I-front signal is approximately 2.64, 0.91, and 0.65 times the total recombination radiation signal, respectively.} Note, however, that the morphology of the emissions are markedly different.  The I-front emission is more extended, spanning $10-20~h^{-1}$cMpc scales, while the recombination radiation is more clumpy. For reference, in Appendix \ref{app:cosmicweb} we provide a map showing only the cosmic web in recombination radiation, i.e. without the neutral islands. An important caveat for Figure \ref{fig:SB_and_LyaRT} is that we do not attempt to model the Ly$\alpha$ from nebular emissions inside of galaxies. LAEs bright enough to be detected could be masked, while the undetected faint LAEs would presumably also appear as peaks clustered outside of the voids -- mostly away from the I-front emissions. Our crude treatment of the recombination contribution to these maps mainly serves to illustrate that the I-front emissions are morphologically different and spatially separated from these sources.   

Lastly, the maps in the bottom row of Figure \ref{fig:SB_and_LyaRT} show the effects of including Ly$\alpha$ RT.  The diffusion of the Ly$\alpha$ photons as they scatter through the IGM smooths out the signal considerably, damping sharp features in the brightness maps. To quantify this smearing scale, we have modified the Ly$\alpha$ RT code to compute the distances of the photons from their sources at the locations of last scattering. We find that $50(90) \%$ of the photons, received at the observer, last scattered within a distance of $0.7(7.5)~h^{-1}$cMpc from their sources, corresponding to an angular scale of $0.44(4.7)$ arcmin in our maps.  We find that the Ly$\alpha$ RT does not have a significant effect on the peak and mean surface brightnesses, with less than 1\% differences in these quantities between the runs with and without Ly$\alpha$ RT. Broadly, we see that the outline of the neutral islands remain intact, as do the peaky structure of the recombination radiation. The recombination radiation features are so sharp to begin with that they remain so even after the effects of Ly$\alpha$ RT are included.  In the next section, we add another layer of realism to these maps by simulating noise for three observational configurations. We will see that the spatially extended nature of the I-front features would, at least in principle, allow them to be distinguished visually under ideal conditions.   

\begin{table*}
    \centering
    \begin{tabular}{c | c | c | c| c}
          Eff. Aperture [m] &  Exp. Time [hr] & Total Eff.  & Sky Rad. [$\frac{\mathrm{ph}}{\mathrm{s}~\mathrm{m}^2 \mu~\mathrm{arcsec}^2}$] & Filter \\
         \hline
        8.2 & 200 & 0.1 & 300 & NB 816\\
        4.5 & 5,000 & 0.1 & 300 & NB 816 
         \end{tabular}
    \caption{Specifications for the observational configurations used in this work. The 4.5m configuration envisions a dedicated instrument designed specifically for low surface brightness observations, and equipped with a filter similar to the NB816 filter on Subaru HSC.}
    \label{tab:scope_table}
\end{table*}

\subsection{Mock Observations} \label{sect:mock_obs}

In this section, we present idealized mock observations to discuss the feasibility of detecting neutral islands with current and futuristic observatories. Again, our focus here is on targeted surveys around quasar sight lines suspected of containing neutral islands owing to the presence of long Ly$\alpha$/Ly$\beta$ forest troughs, e.g. J0148 \cite{2015MNRAS.447.3402B}. For each pixel we compute a photon count using, 
\begin{equation}
N_{\mIF} = \dot N_{\mIF} t_\mathrm{exp} = T\frac{\mathrm{SB}_{\mlya}}{h\nu_{\mlya}} \pi \left(\frac{D}{2}\right)^2 \Omega_\mathrm{pixel}t_\mathrm{exp},
\end{equation}
where $\dot N_{\mIF}$ is the count rate per pixel, $T$ is the total efficiency  (combined effects from the sky, telescope, and instrument),
$\mathrm{SB}_{\mlya}$ is the \lya\ surface brightness of the source, $D$ is the aperture diameter, $\Omega_\mathrm{pixel}$ is the pixel scale, and $t_{\rm exp}$ is the exposure time. 

We assume that the noise is completely sky dominated. The average number of sky photons detected per pixel is
\begin{equation}
    N^{\mathrm{sky}} = T \frac{I_\lambda^\mathrm{sky} \Delta\lambda}{h\nu_{\mlya}}  \pi \left(\frac{D}{2}\right)^2 \Omega_\mathrm{pixel} t_\mathrm{exp}
\end{equation}
where the sky radiance, $I_\lambda^{\rm sky}$, has units of [erg s$^{-1}$ cm$^{-2}$ \AA$^{-1}$ arcsec$^{-2}$]. We randomly draw the number of sky photons detected in each pixel from a Poisson distribution with average $N^{\mathrm{sky}}$. We then subtract a constant $ N^{\mathrm{sky}}$ from each pixel, assuming optimal sky subtraction.  Sky subtraction and foreground removal would clearly be crucial factors for achieving the desired sensitivity in real observations.
Our purpose here is to assess what is in principle possible under ideal assumptions.   

\begin{figure*}
    \centering
    \includegraphics[width=1.0\textwidth]{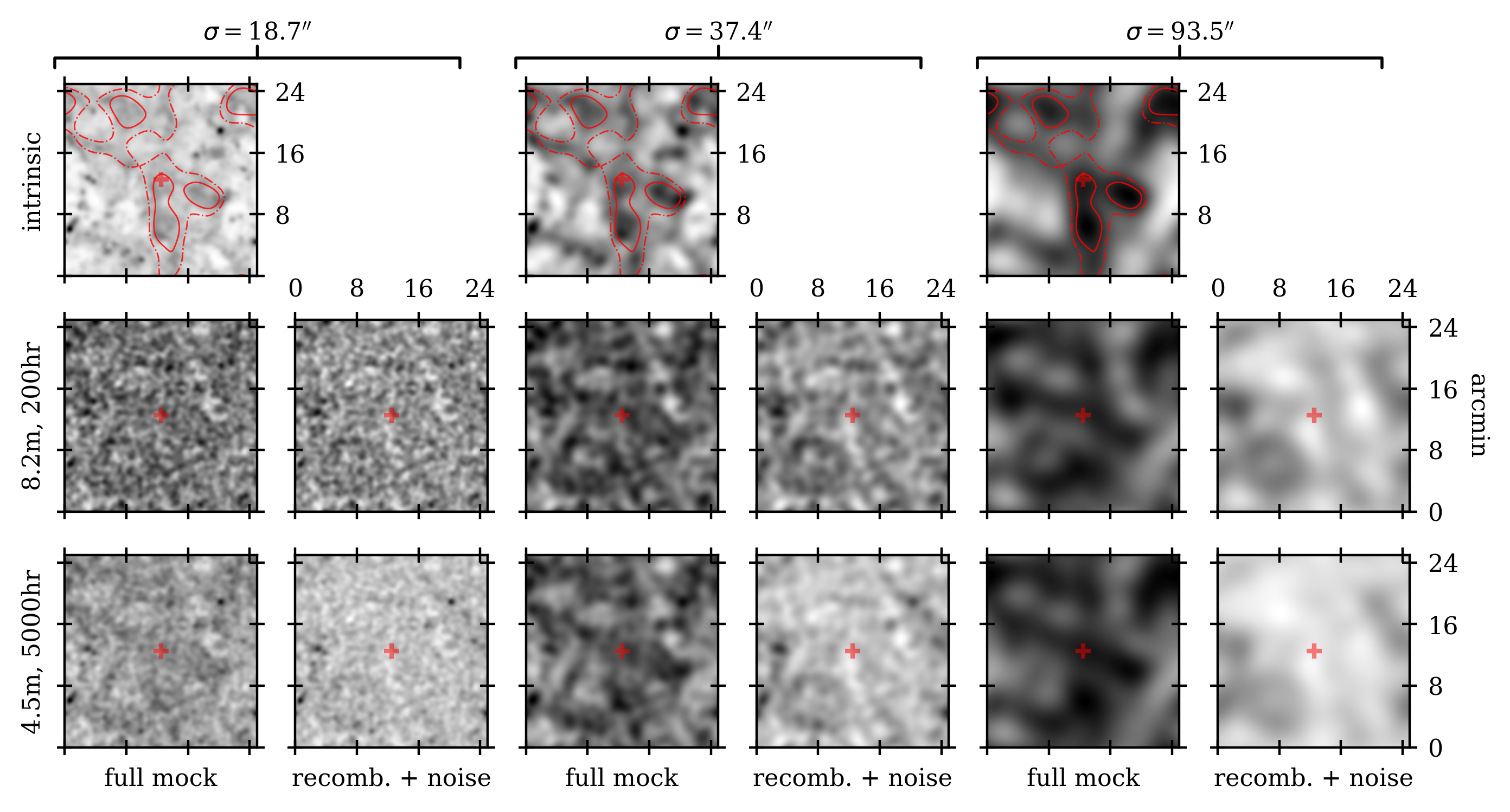}
    \caption{Mock narrowband images of neutral islands near the end of reionization. Darker shading corresponds to brighter emission. In this figure we adopt the reionization model, $\alpha=0.5,$ \textsc{faster}-$v_\mathrm{IF}$ (see main text), which yields brighter I-front emissions owing to the hard spectrum and fast-moving I-fronts.  We smooth the images using a Gaussian kernel with standard deviation, $\sigma$, annotated at the top of the figure.  The columns labeled ``full mock'' include signals from neutral islands and recombination radiation, and they include sky noise.  The top row, labeled ``intrinsic'', shows the images effectively for an infinite integration time. To aid the visual identification of neutral islands in the maps, the red contours in those panels show the 70th and 90th percentiles of the I-front surface brightness for the case with $\sigma = 93.5''$. The 2nd and 3rd rows correspond to the 2 observational setups summarized in Table \ref{tab:scope_table}. For comparison, each full mock column is displayed next to a column for which the neutral islands are removed (these are labeled recomb.+noise). This comparison mimics an observational setup where an image of a quasar-centered field containing neutral islands is compared against random fields elsewhere. All images in the 2nd and 3rd rows assume optimal background subtraction and foreground removal. Smoothing the images on larger scales makes it possible to distinguish a field with neutral islands from one without them.}    
    \label{FIG:MOCK_MAPS}
\end{figure*}

\begin{figure*}[h]
    \centering
    \includegraphics[width=0.8\textwidth]{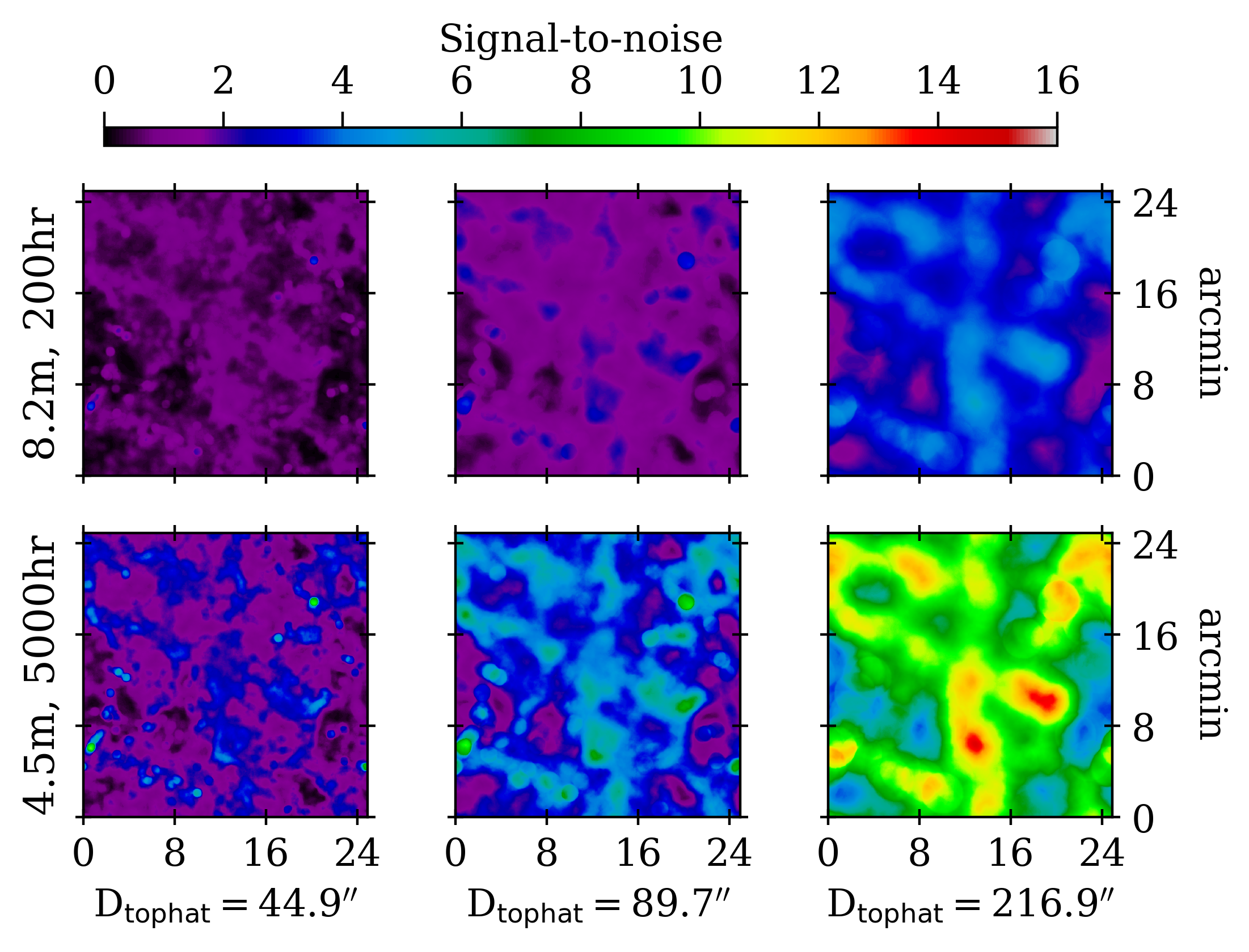}
    \caption{Signal-to-noise (S/N) maps for the three observational scenarios in Figure \ref{FIG:MOCK_MAPS}. For each pixel, the signal maps are summed within a circular, top-hat aperture with a diameter equal to the FWHM of the Gaussian kernels used in Figure \ref{FIG:MOCK_MAPS}. The corresponding aperture diameters are given on the bottom axes of the panels.  The noise is taken to be $\sqrt{N_\mathrm{pix}} \sigma^{\rm noise}_\mathrm{pix}$, where $N_\mathrm{pix}$ is the number of pixels within the aperture and $\sigma^{\rm noise}_\mathrm{pix}$ is the standard deviation of the un-smoothed noise distribution. These S/N maps correspond to the 6 full mock images in Figure \ref{FIG:MOCK_MAPS}.
    }
    \label{fig:SNR}
\end{figure*}

We explore two observational scenarios summarized in Table \ref{tab:scope_table}. Both scenarios assume a narrowband filter with characteristics similar to the NB816 filter on Subaru Hyper Suprime-Cam (targeting \lya\ at $z=5.726$). The first scenario adopts an aperture of 8.2m, similar to Subaru. We use an ambitious integration time of $200$ hours, roughly forty times longer than the NB816 imaging toward J0148 previously carried out \cite{2023ApJ...955..138C}. Given the spatially extended nature and low surface brightness of the I-front emissions, it is useful to consider an alternative setup specifically designed to minimize the background from scattering within the telescope -- a chief impediment in low surface brightness observations. To this end, our second scenario is modeled on the Dragonfly telephoto array \cite{2014PASP..126...55A,2020ApJ...894..119D}.   While the current configuration with 48 lenses has too small of an aperture for our purposes (effectively 0.99 m), we consider the hypothetical 1,000-lens 4.5m effective aperture configuration as in Ref. \cite{2019ApJ...877....4L} with a long integration time of 5,000 hours. This setup models a scenario in which a dedicated instrument is built for the purpose of detecting I-fronts during reionization. For both scenarios we adopt a total sky+telescope+instrument throughput of $10\%$, and a sky radiance of $300$ photons s$^{-1}$m$^{-2}\mu^{-1}$arcsec$^{-2}$, reasonable for a new moon near 8160 \AA.\footnote{The sky radiance is calculated from the Cerro Paranal mountain in Chile with a new moon and an airmass of 1 near 8160 \AA. The relevant sky radiances should be reasonably similar at other locations. See \href{https://www.eso.org/observing/etc/bin/gen/form?INS.MODE=swspectr+INS.NAME=SKYCALC}{Cerro Paranal sky calculator web application.}}  Most of our results use a field of view of $24.9'\times24.9'$, matched to the $40\times40$ ($h^{-1}$cMpc)$^2$ area provided by our simulations, but we will discuss larger fields at the end of this section. It is worth noting that the larger fields-of-view that would be obtained in practice for both observational setups in Table \ref{tab:scope_table} could provide additional advantages in identifying neutral islands visually, since the islands are expected to be highly clustered in cosmological voids (about this more below).  We also explore different smoothing scales for our mock images in an attempt to exploit the large spatial extent of the I-front emissions. For our mock images, we apply a Gaussian smoothing kernel with standard deviation, $\sigma$.

Figure \ref{FIG:MOCK_MAPS} shows mock observations for our most optimistic model, \textsc{$\alpha = 0.5$, faster-$v_\mathrm{IF}$}.  All panels correspond to the same view -- i.e. the same region of our simulation volume -- as in Figure \ref{fig:SB_and_LyaRT}. Darker shading corresponds to brighter emission. For reference, the top row provides the intrinsic signal, effectively for an infinite integration time. To aid the visual identification of neutral islands in the maps, the red contours in those panels show the 70th and 90th percentiles of the I-front surface brightness for the case with $\sigma = 93.5''$. As noted by the brackets above, the 1st-two, 2nd-two, and 3rd-two columns correspond respectively to smoothing scales of $\sigma = 18.7''$, $37.4''$, and $93.5''$. At $z=5.7$, these translate to transverse distances of 0.5, 1 and 2.5 $h^{-1}$cMpc, 
respectively. The 2nd and 3rd rows correspond to the two observational setups as noted on the left-most axes.    The columns labeled ``full mock'' include signals from neutral islands and recombination radiation, both processed with Ly$\alpha$ RT, as well as the effects of sky noise according to the observational configuration.  The columns labeled ``recomb.+noise'' exclude the neutral island signal. Each pair of full mock and recomb+noise panels uses the same color scale. For instance, the panel in the 2nd row, 1st column shares the same color scale as the panel directly to its right. For a fixed observational setup and smoothing scale, these pairs compare a field in which there are neutral islands (full mock) to the same field without them (recomb.+noise). This comparison mimics an observational setup where a quasar-centered image is compared against random fields elsewhere, less likely to contain neutral islands. The question at hand is: can the former field containing the neutral islands be distinguished from fields that do not?

Figure \ref{fig:SNR} provides corresponding signal-to-noise ratio (S/N) maps for the mock images in Figure \ref{FIG:MOCK_MAPS}. To obtain these, we first convolve a signal map with a circular, top-hat aperture with diameter equal to the full width at half maximum (FWHM) of the Gaussian kernel used in Figure \ref{FIG:MOCK_MAPS}. For reference, the FWHMs for $\sigma = 18.7''$, $37.4''$, and $93.5''$ are $44.9''$, $89.7''$, and $216.9''$, respectively.  The noise is taken to be $\sqrt{N_\mathrm{pix}} \sigma^{\rm noise}_\mathrm{pix}$, where $N_\mathrm{pix}$ is the number of pixels within the aperture and $\sigma^{\rm noise}_\mathrm{pix}$ is the standard deviation of the un-smoothed noise distribution. The sharp circular features seen in some locations of the maps are a result of the top-hat filter passing narrow peaks in the signal from recombination radiation. 

Smoothing the images on larger scales makes it possible to distinguish a field with neutral islands from one without them.  For instance, consider the 2nd row in Figure \ref{FIG:MOCK_MAPS} for the (8.2m, 200hr) setup. The neutral island signal is difficult to discern in the left-two panels, which correspond to the smallest smoothing scale. Moving rightward, i.e. toward larger smoothing scales, it becomes increasingly obvious that there is a neutral island signal. This is seen in the 1st row of Figure \ref{fig:SNR} as well. 
 The S/N of the neutral island features is $\approx 1.7$ for the smallest smoothing scale, but reaches $\approx 5$ in the right-most map. Naturally the situation is improved for the (4.5m, 5,000hr) setup that is shown in the bottom rows of both figures. In this case, the neutral island features reach S/N $\approx 14$.   An important caveat is that we have assumed optimal background subtraction and foreground removal. In reality, uncertainties in the zero point may make it difficult to discern neutral islands from the highly smoothed features originating from recombination radiation or unresolved LAEs (more about the latter below).  For example, the recombination radiation feature at $(x,y) = (4',8')$ in the bottom-right image of Figure \ref{FIG:MOCK_MAPS} (which does not contain neutral islands) could be confused for a neutral island if the absolute scale of the image were not known. In this case, there is a tradeoff between image resolution and S/N. On the other hand, an image processing approach that is more tailored than the simple smoothing used here could allow the sharp peaks from recombinations to be distinguished from the diffuse large-scale emissions from neutral islands.

\begin{figure}
    \centering
    \includegraphics[width=1.0\textwidth]{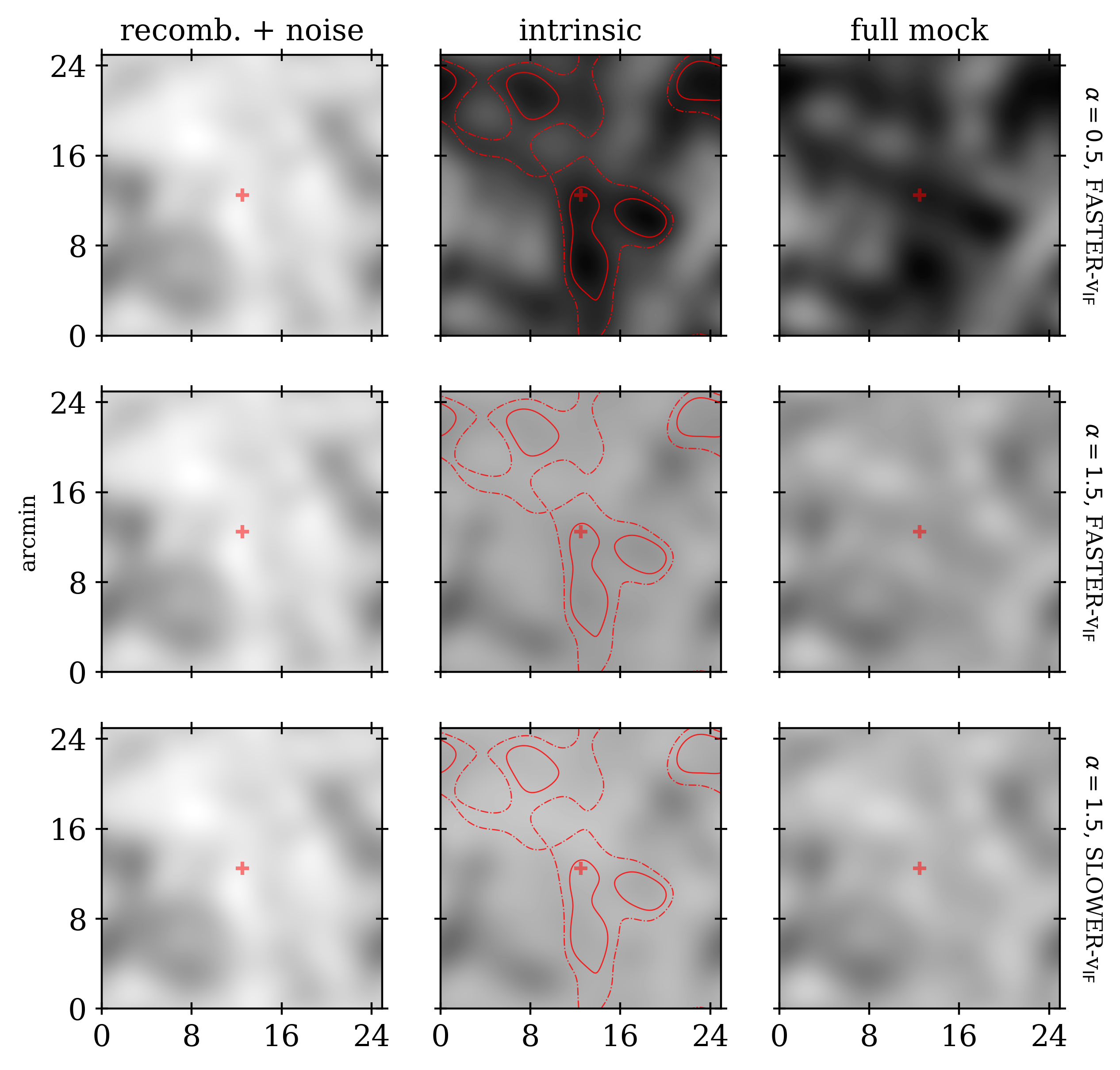}
    \caption{Same as Figure \ref{FIG:MOCK_MAPS}, but now additionally including the $\alpha=1.5$, \textsc{faster}-$v_\mathrm{IF}$ and $\alpha=1.5$, \textsc{slower}-$v_\mathrm{IF}$ models. All panels assume the (4.5m, 5,000hr) setup, adopt a smoothing kernel with $\sigma = 93.5''$, match the color scale of the rightmost two columns of Figure \ref{FIG:MOCK_MAPS}, and are centered on the same neutral island as the previous figures. For reference, the top row shows the more optimistic $\alpha=0.5$, \textsc{faster}-$v_\mathrm{IF}$ model. The dark blobs in the second and third row of the intrinsic signal column are from recombination radiation while the faint gray features near the center of the maps are from \IF\ emissions. The recombination radiation is significantly brighter than the I-fronts bounding the neutral islands in the second and third rows. Distinguishing the extended I-front emission from the more compact (and, in this case, brighter) recombination radiation would require more sophisticated image processing than what is presented here.
    }
    \label{fig:other2_models}
\end{figure}

The discussion so far has been centered on our most optimistic reionization model. The signal is substantially dimmer in the region of parameter space where the ionizing spectrum is softer and/or the I-fronts move slower, i.e. where reionization ends more gradually.  Figure \ref{fig:other2_models} compares mock observations for all three models: \textsc{$\alpha = 0.5$, faster-$v_\mathrm{IF}$}; \textsc{$\alpha = 1.5$, faster-$v_\mathrm{IF}$}; and \textsc{$\alpha = 1.5$, slower-$v_\mathrm{IF}$}. All panels adopt the most favorable configuration of $D=4.5\mathrm{m}$, $t_{\mathrm{exp}}=$ 5,000hr and $\sigma=93.5''$.  In the \textsc{$\alpha = 1.5$, faster-$v_\mathrm{IF}$} and \textsc{$\alpha = 1.5$, slower-$v_\mathrm{IF}$} models, the signal from recombination radiation is significantly brighter than that of the neutral islands. The dark features seen on the left and right sides of the intrinsic signal panels are from recombination radiation, while the fainter features near the centers of the maps are from \IF\ emissions.  Comparing the middle and right columns, we can nonetheless discern the fields containing the neutral islands under our idealized assumptions.  As noted above, distinguishing the extended I-front emission from the more compact (and, in this case, brighter) recombination radiation would require more sophisticated image processing than what is presented here. 

\begin{figure*}
    \centering
\includegraphics[width=\textwidth]{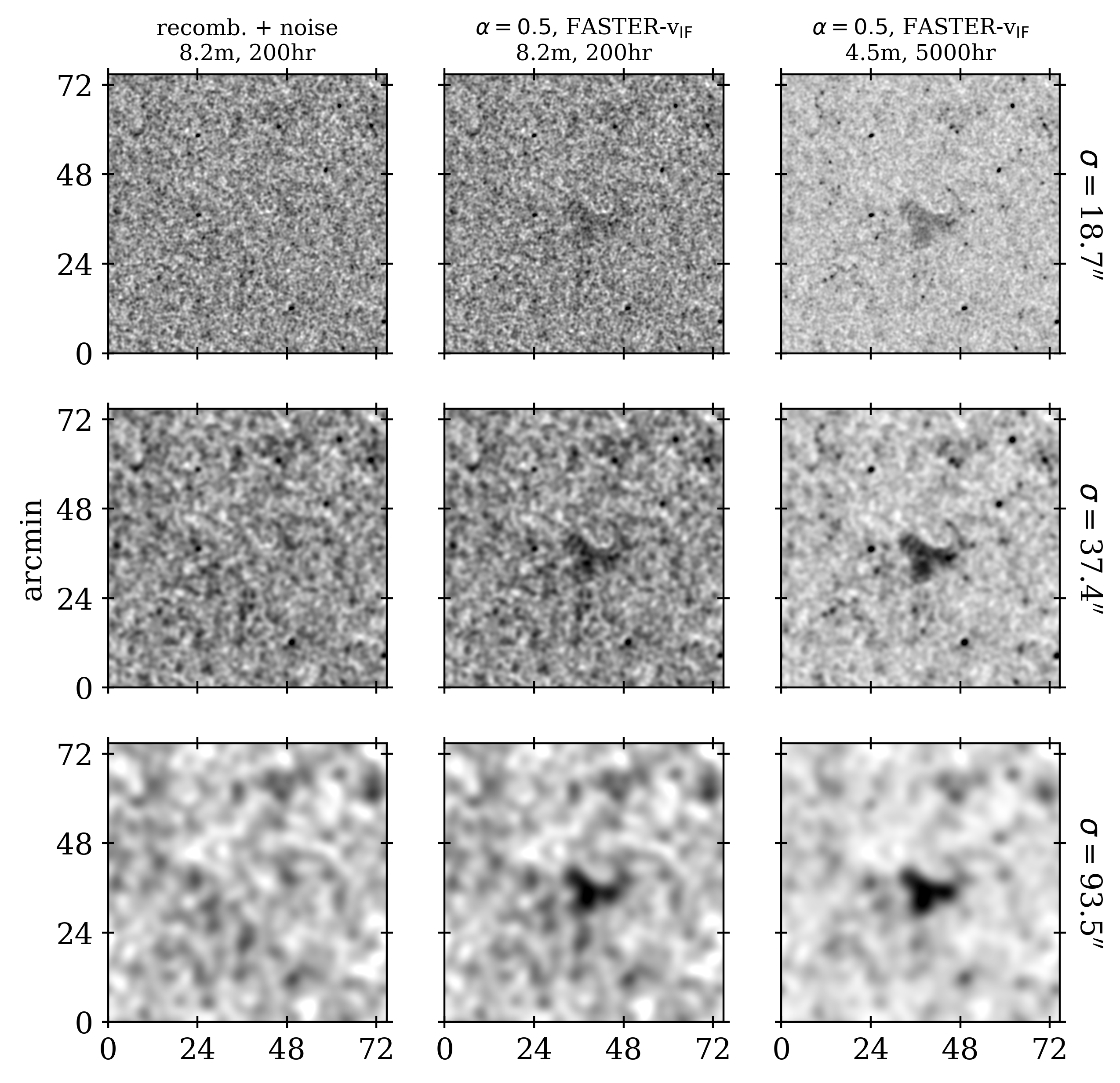}
    \caption{The effect of a wider field of view for neutral island visibility. Here we construct $72'$ by $72'$ fields of view (compare against the $24'$ by $24'$ fields in Figure \ref{FIG:MOCK_MAPS}). We place a neutral island from the $\alpha = 0.5$, \textsc{faster}-$v_{\rm IF}$ model at the center of the map and create a wider field by tiling around it with random slices of a simulation that includes only recombination radiation. The isolated large-scale emission from the neutral island makes it significantly easier to pick out of a wider field when the map is smoothed on large scales.}
    \label{fig:wide_map}
\end{figure*}

The mock images shown thus far were constructed from reionization simulations with relatively small volumes ($L= 40~h^{-1}$cMpc), with neutral islands spanning much of the box (see bottom-left panel of Figure \ref{fig:SB_and_LyaRT}). This has limited us to focusing mostly on the contrast between fields with and without neutral islands. However, the fields of view of actual wide-field imagers would be substantially larger than we are able to simulate here; the neutral islands would likely be clustered in large-scale voids that our simulations do not sample.  This could result in image geometries where there is greater contrast between regions with and without I-front emission. To illustrate this effect, in Figure \ref{fig:wide_map} we crudely construct a wider field of view of $72'$ by $72'$ from our simulations. We manually select out the triangular neutral island feature with top vertex at the center of the field, as seen in second panel in the bottom row of Figure \ref{fig:SB_and_LyaRT}. Note that this includes the effects of Ly$\alpha$ RT.  We place this region at the center of the map. We then tile around the neutral island by randomly slicing regions from simulations that include emissions from recombination radiation only. Since the maps are convolved with a relatively wide Gaussian kernel in the end, we do not include the effects of Ly$\alpha$ RT for the recombination radiation-only tiles. Finally, we add noise to the widened mock images as before.  

The panels in Figure \ref{fig:wide_map} show the results of this exercise for the \textsc{$\alpha = 0.5$, faster-$v_\mathrm{IF}$} model. The rows show different smoothing kernel widths.  The 2nd and 3rd columns correspond to the (8.2m, 200hr) and (4.5m, 5,000hr) setups, respectively. Each panel uses the same color scale as its analogue in Figure \ref{FIG:MOCK_MAPS}. For comparison, the first column shows the case with no neutral island for the (8.2m, 200hr) setup; it includes only the recombination radiation and noise. This column can be compared directly to the 2nd column.    The large-scale contrast that results from the isolated neutral island at the center of the map makes it significantly more visible when the map is smoothed on larger scales. The situation could thus turn out to be more optimistic than our main results (based on smaller fields) suggest, motivating further development of the models. Figure \ref{fig:wide_map_other_models} in Appendix \ref{appendix:othermodels} shows wider-field images for the \textsc{$\alpha = 1.5$, faster-$v_\mathrm{IF}$} and \textsc{$\alpha = 1.5$, slower-$v_\mathrm{IF}$} models. Even in these cases, for which the signal is much weaker, the neutral island is visible.  As anticipated, however, without prior knowledge about their locations, the neutral islands are much more difficult to tell apart from the recombination radiation. 

In summary, our results indicate that directly imaging the I-front emissions will be challenging but potentially within reach of an ambitious observing program. It would require optimal sky subtraction and excellent mitigation of scattered light, and a favorable reionization scenario resulting in the neutral islands being particularly bright: hard ionizing spectra and fast-moving I-fronts.  We caution, however, that the maps in Figure \ref{fig:wide_map} are idealized in two important ways: (1) we have artificially placed a single neutral island at the center of the map.  However, in a field of view of $72'$ by $72'$, there would likely be other neutral islands that could make the signal toward the quasar sight line less conspicuous; (2) we have not included emissions from LAEs, which could also act to make the signal less conspicuous.  Of course, bright LAEs detected in the field would be masked. Potentially more problematic are the faint LAEs whose clustered intensity, if bright enough, could be confused for neutral island emission. We estimate the magnitude of this effect using empirical measurements of the Ly$\alpha$ luminosity function by Ref. \cite{2018PASJ...70S..16K}. Consider the (8.2m, 200hr) setup.  The luminosity limit of Ref. \cite{2018PASJ...70S..16K} is around $\log(L_{\rm{Ly}\alpha}/[\rm{ergs~s}^{-1}]) = 42.4$ for an exposure time $t_{\rm exp} = 5.5$ \cite{2018PASJ...70S..14S}.  The corresponding limit for 200 hours of integration would be $\log(L_{\rm{Ly}\alpha}/[\rm{ergs~s}^{-1}]) \sim 41.6$, assuming a scaling of $1/\sqrt{t_{\rm exp}}$. The behavior of the faint end of the luminosity function, below current limits, is highly uncertain, but we generally expect a flattening and cutoff below some unknown luminosity.  For guidance on where to place this cutoff, we invoke the model of Ref. \cite{2018PASJ...70...55I} connecting Ly$\alpha$ luminosity to halo mass, which is motivated by radiation hydrodynamics simulations (see their eq. 6). Their model suggests that a halo mass of $\sim 10^{9}$M$_{\odot}$ corresponds roughly to $\log(L_{\rm{Ly}\alpha}/[\rm{ergs~s}^{-1}]) \sim 40.7$, on average. Since star formation is generally expected to be less efficient in halos of lower mass, owing to feedback processes \cite{2014MNRAS.443L..44B,2014MNRAS.444..503N}, we will adopt this lower limit. Under these assumptions, we estimate a surface brightness for unresolved LAEs of $\sim 7\times 10^{-21}$ erg s$^{-1}$ cm$^{-2}$ arcsec$^{-2}$. This is similar to the brighter limit of I-front emissions in our models, but our estimate here is based on the average luminosity function. The targets of interest for imaging neutral islands are known to be large-scale voids (e.g. the sight line of J0148), so the local LAE surface brightness may be considerably lower (see e.g. Ref. \cite{2021MNRAS.501.5294G}). Nonetheless, this is an topic that must be explored in greater detail. We plan to address both of the above caveat in future work by applying our modeling framework to reionization simulations of larger volume.

\section{Conclusion}\label{sect:conclusion}

Observational evidence and theoretical modeling strongly suggest that the long troughs observed in the $z > 5.5$ Ly$\alpha$ and Ly$\beta$ forests arise from neutral islands during the last phases of reionization.  If this is true, then the longest such troughs tell us exactly where to witness reionization in progress.  In this paper we have explored the possibility of imaging the reionizing IGM directly using Ly$\alpha$ emissions from the I-fronts bounding neutral islands.  The obvious first candidate for such an observation would be the sight line toward quasar J0148, which has been discussed extensively in the literature owing to the $\approx 110~h^{-1}$cMpc Ly$\alpha$ trough in its spectrum. 

In Paper I we used high-resolution one-dimensional RT simulations to understand the parameter space of I-front Ly$\alpha$ emissions. Paper I makes public a convenient interpolation table for the I-front Ly$\alpha$ production efficiency, which can be used as a sub-grid model in larger-volume reionization simulations that do not resolve the emission processes inside of I-fronts.  Here, we have run cosmological RT simulations of reionization to model neutral islands at $z\approx 5.7$, like the ones that are hypothesized to inhabit the J0148 sight line.  We applied the results from Paper I to model the Ly$\alpha$ emissions from these neutral islands.  Our models span a range of plausible spectral indices for the ionizing radiation background and I-front speeds -- the main physical parameters determining the brightness of the Ly$\alpha$ emissions.  We have also included the effects of Ly$\alpha$ RT.   

We found that the I-fronts bounding neutral islands emit at brightnesses $\approx 1 - 5\times 10^{-21}$ erg s$^{-1}$ cm$^{-2}$ arcsec$^{-2}$.  The lower limit of this range corresponds to a model with spectral index of incident ionizing radiation $\alpha = 1.5$, and an average I-front speed of $v_{\rm IF}\approx 5\times 10^{3}$ km s$^{-1}$.   The upper limit reflects a model with $\alpha = 0.5$ and average $v_{\rm IF}\approx 10^{4}$ km s$^{-1}$.  The hardness of the ionizing radiation in the latter model is motivated by our lack of knowledge about the sources that drove reionization, as well as the possibility of spectral filtering by the IGM if much of the radiation has to travel $\sim 10~h^{-1}$cMpc from its sources.  The upper range of the brightnesses quoted above is similar to those expected from recombination radiation in the density peaks of the cosmic web.  The expected I-front signal, however, is geometrically distinct, consisting of large-scale features extending over tens of cMpc/$h$ (tens of arcminutes).
For the most part, it is also separated spatially from the recombination emission because the neutral islands are isolated to cosmic voids near the end of reionization.

We then considered the prospects for directly imaging the neutral islands in Ly$\alpha$. We developed mock observations for two scenarios.  The first assumes a narrowband imaging campaign of 200 hours on a telescope with 8.2m aperture. The second is modeled after a futuristic instrument with 4.5m aperture specifically designed for low surface brightness observations, minimizing the effects of scattered light to improve sky background subtraction. One possibility for the latter is a future Dragonfly-like telephoto array with appropriate filters installed. We found that neutral islands may be detectable with an ambitious observing program like our two model scenarios, especially if the I-fronts are on the brighter side of expectations, e.g. our \textsc{$\alpha = 0.5$, faster-$v_\mathrm{IF}$} model.  Smoothing the image on large scales $\gtrsim 2~h^{-1}$cMpc helps amplify the large-scale signal of the neutral islands over the noise and smaller-scale knots from recombination radiation. The prospects are better if the neutral island is isolated at the center of a wide field of view, as illustrated in Figure \ref{fig:wide_map}. This may be the case in the sight line toward J0148, for example. One caveat is that our mock observations assume optimal sky subtraction and foreground removal, including excellent mitigation of scattering light. It may also require the modeling and subtraction of unresolved light from faint LAEs below the detection threshold of the survey, which may be partly mitigated if the I-front emission is located in rare voids. Lastly, the neutral islands would be more difficult to detect if the I-front emissions are on the dimmer side of expectations, from softer incident ionizing radiation and/or slower moving I-fronts. A corollary of these results is that a detection of the neutral islands could place limits on the hardness of the ionizing radiation background driving the I-fronts. 

A natural followup question is whether a dedicated space-based observatory could significantly improve the prospects. Observing from space would decrease the sky background substantially.  However, a competing consideration is the relatively large aperture sizes that would be needed to achieve the required sensitivity. Narrower filters could potentially help in wide-field observations from the ground by cutting down the sky foreground, but would require the fortuitous placement of a neutral island within the bandwidth.

Imaging neutral islands by way of their I-front Ly$\alpha$ emissions would give us a direct view of a reionizing IGM.  Such images would not only constrain the timing of reionization, they could also help elucidate the connection between the high-$z$ galaxy population and their cosmological environments.  The technology required for such an endeavor would have many other impactful applications.  For instance, imaging the Ly$\alpha$ emissions of the cosmic web, or revealing ultra-faint LAEs, would both be of great interest to the broader cosmology community.  Our results help motivate the continued development of low surface brightness observation methods. 

\acknowledgments
We thank Matt McQuinn for helpful comments on this manuscript, Hy Trac for providing the cosmological hydrodynamics code that was used in this work, and Zheng Zheng for helpful discussions. A.D.’s group was supported by NASA 19-ATP19-0191, NSF AST-2045600, and JWSTAR-02608.001-A. E.V. is supported by NSF grant AST-2009309 and NASA ATP grant 80NSSC22K0629. Some of the results in this paper have been derived using the HEALPix \cite{2005ApJ...622..759G} package. Some of this work was carried out at the Advanced Research Computing at Hopkins (ARCH) core facility  (rockfish.jhu.edu), which is supported by the National Science Foundation (NSF) grant number OAC1920103. Expanse is an NSF-funded system operated by the San Diego Supercomputer Center at UC San Diego, and is available through the ACCESS program.

\bibliography{references.bib}
\bibliographystyle{JHEP}

\appendix

\section{Spectral hardening}\label{appendix:hardening}

Paper I shows that the spectral shape of the incident ionizing spectrum plays a significant role in the amount of \lya\ photons emitted from I-fronts. In addition to the properties of the sources, the incident spectral index depends on the amount of filtering by H absorption as the radiation traverses long distances through the IGM.   For the main results of this paper, we use a monochromatic RT approach to reduce computational cost, exploring instead a range of constant incident spectral indices.  In this appendix, we provide some calculations to estimate the expected amount of spectral filtering by the IGM. 

 \begin{figure}
    \centering
\includegraphics[width=0.45\textwidth]{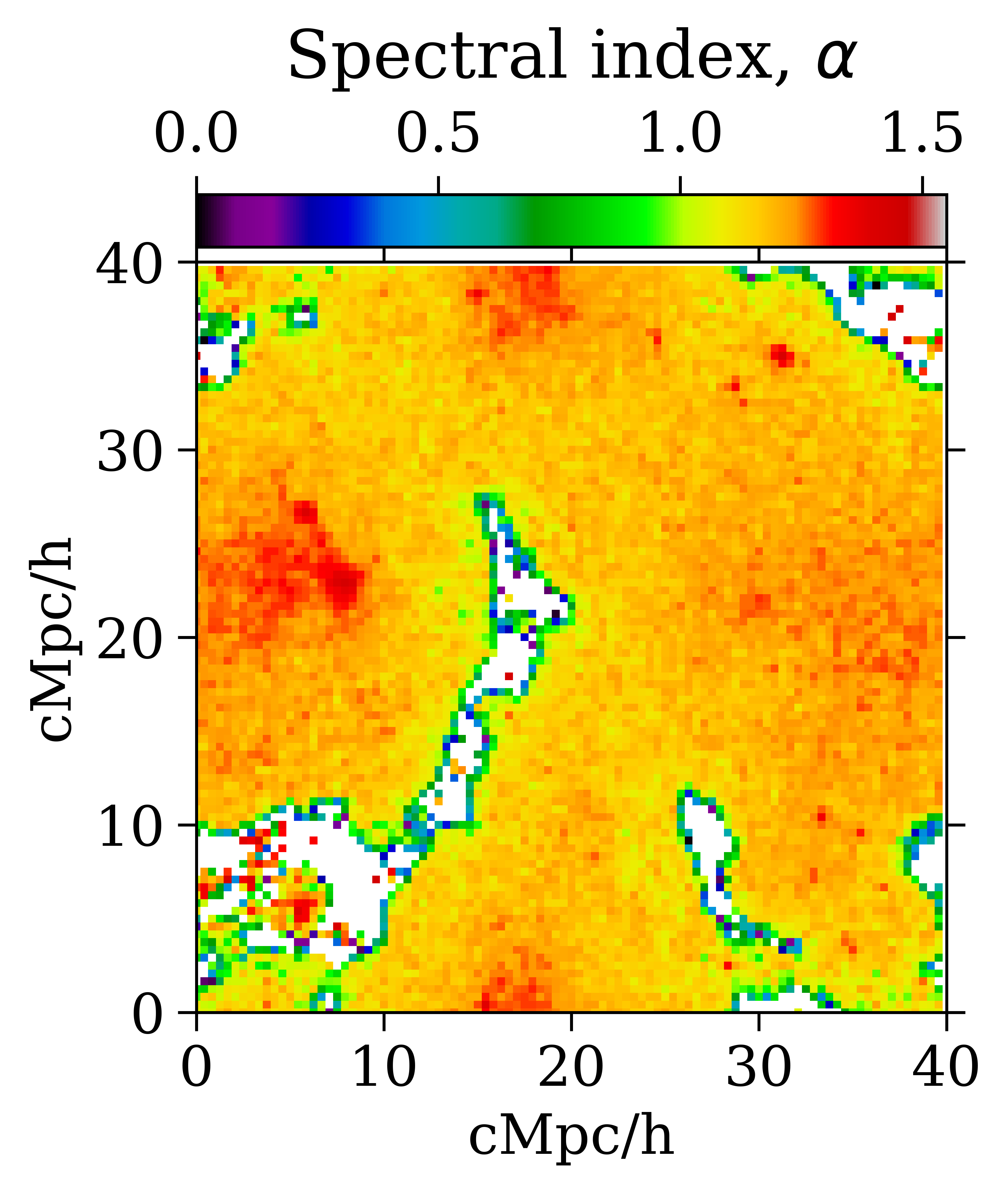}
    \caption{Effective spectral index map in a multi-frequency RT simulation of reionization with $N=96^3$ RT cells. Residual neutral gas preferentially filters out ionizing radiation closer to 1 Ryd resulting in spectral hardening. The spectral index of the radiation field near the sources is similar to that of the sources (red regions in the map).  The spectrum of the background becomes harder at further distances from the clustered sources. Towards the end of reionization, radiation travels across longer distances through the IGM to ionize the neutral islands, which are relegated to the voids in the density field.  This results in harder spectra for the radiation driving the I-fronts bounding these neutral islands.}
    \label{fig:z5.83_alpha}
\end{figure}

The calculation that we offer here compliments that in the appendix of Paper I, which is based on a suite of high resolution 1D-RT simulations.  Here, we use a lower resolution $N=96^3$ multi-frequency FlexRT simulation of reionization (in 3D) to estimate the degree of spectral hardening. (Note that the main runs used in this paper are monochromatic with $N=400^3$.) The simulation was run in the same $L=40~h^{-1}$cMpc box used for our main results. The five frequency bins are centered on $[14.48,16.70,20.03,25.78,39.23]$ eV, chosen so there are initially an equal fraction of photons in each bin for a source spectrum of $\alpha_* = 1.5$. Similar to the calculations in the appendix of Paper I, the mean photoionization rate of $\Gamma_{\rm HI} = 0.262\times 10^{-12}$ s$^{-1}$ was calibrated to approximately match the measurements of Ref. \cite{2023MNRAS.525.4093G}.

 In Figure \ref{fig:z5.83_alpha} we show a map of spatial variations in the effective spectra index measured from the simulation at a neutral fraction of $22\%$. The slice shown is one-cell thick; it is roughly the same slice as in Figure \ref{fig:gas_slices}, but with lower RT resolution.  As in our main runs, the source spectral index is set to be $\alpha_{*}=1.5$. Regions closer to ionizing sources have effective spectral indices closer to $\alpha_{*} = 1.5$.  Towards the end of reionization, radiation travels across longer distances through the IGM to ionize the neutral islands, which are relegated to the voids in the density field.  This results in harder spectra for the radiation driving the I-fronts bounding these neutral islands.  For example, at distances $\sim 10~h^{-1}$cMpc away from the sources, the spectrum hardens by $\Delta \alpha = \alpha_{*}-\alpha \approx 0.5$, in approximate agreement with the Paper I calculation using 1D-RT. According to the results of Paper I, this could correspond to a boost in the I-front \lya\ production efficiency of $\Delta \zeta \approx 50\%$.  It is important to note that the self-shielding of small-scale density peaks in the IGM, which likely play a significant role in the spectral hardening, is not resolved in this low resolution simulation.  With this limitation in mind, these results should be seen as complimentary to the high-resolution calculations presented in the appendix of Paper I. The results presented here and in Paper I, along with the large uncertainties in the source spectra, motivate the wide range of $\alpha = 1.5$ to $\alpha = 0.5$ explored in the main results of this paper. 

\section{Surface brightness map of the cosmic web without neutral islands}
\label{app:cosmicweb}

\begin{figure}
    \centering
\includegraphics[width=0.45\textwidth]{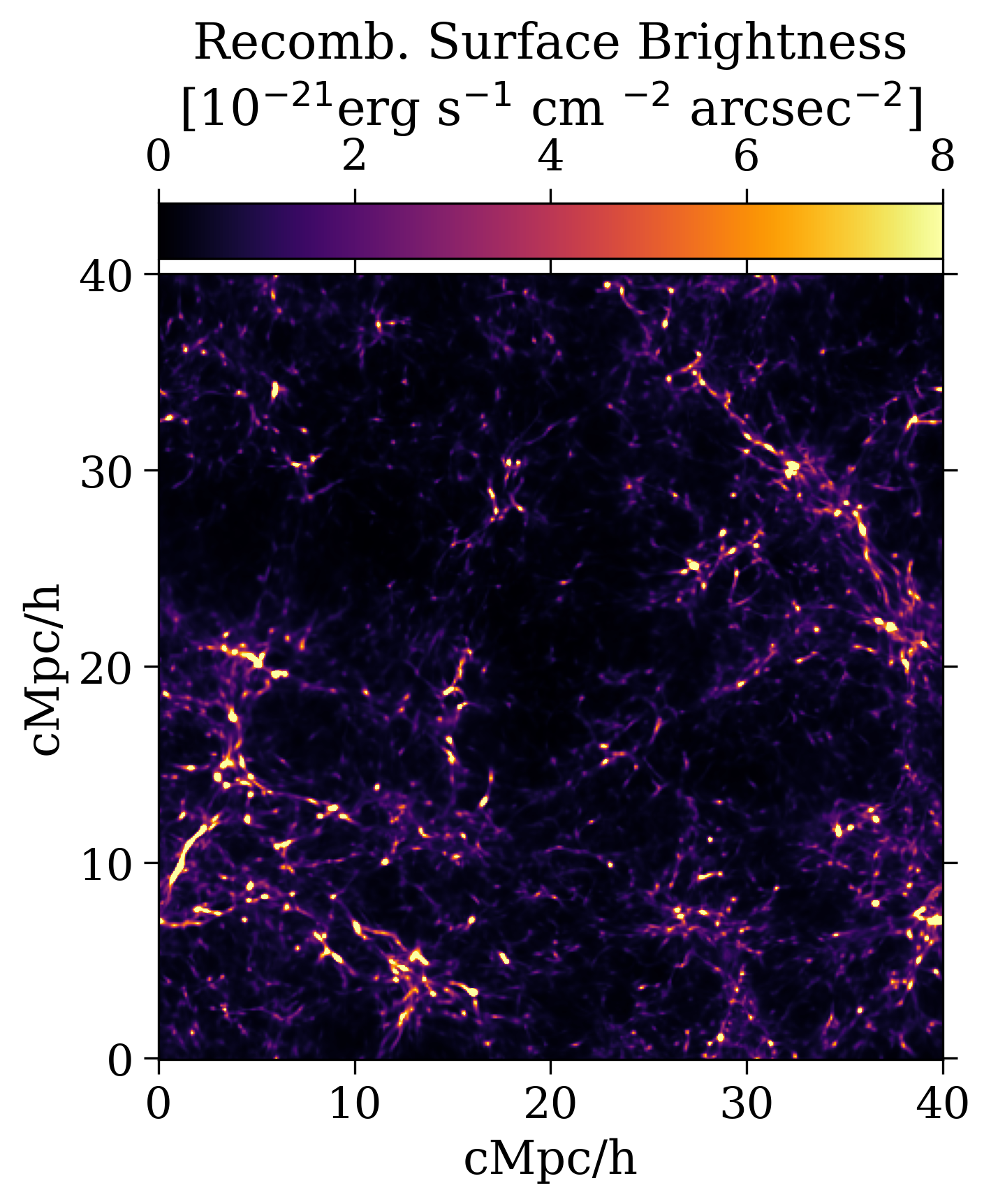}
    \caption{Same as the surface brightness panels of Figure \ref{fig:SB_and_LyaRT}, except that this map contains only the recombination radiation from the cosmic web (without \lya\ RT). This panel may be compared directly to the top two rows of Figure \ref{fig:SB_and_LyaRT}.}
    \label{fig:recomb_only}
\end{figure}

In Figure \ref{fig:recomb_only}, we show a surface brightness map that includes only the contribution of recombination radiation from the cosmic web. This panel may be compared directly to the top two rows of Figure \ref{fig:SB_and_LyaRT}, which also contain emissions from neutral islands. The recombination signal is most intense in compact, high-density nodes of the cosmic web, which contrasts with the more diffuse \IF\ radiation from neutral islands in the voids.

\section{Wider-field images in the $\alpha=1.5,$ \textsc{faster}-$v_\mathrm{IF}$ and $\alpha=1.5,$ \textsc{slower}-$v_\mathrm{IF}$ models}

\label{appendix:othermodels}

\begin{figure*}
    \centering
\includegraphics[width=\textwidth]{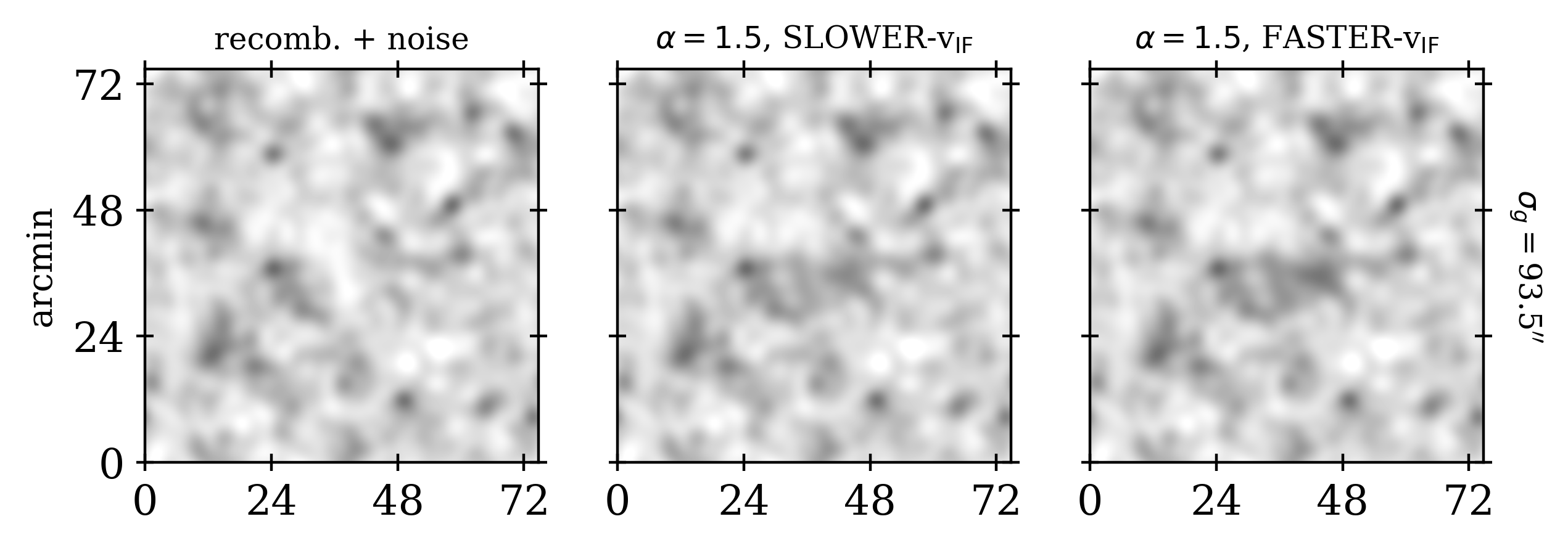}
    \caption{The effect of a wider field of view for neutral island visibility in the \textsc{$\alpha = 1.5$, slower-$v_\mathrm{IF}$} and \textsc{$\alpha = 1.5$, faster-$v_\mathrm{IF}$} models. We construct $72'$ by $72'$ fields in the same manner as to those in Figure \ref{fig:wide_map}. Here, we assume the most favorable configuration of $D=4.5$, $t_\mathrm{exp}=$ 5,000, and $\sigma=93.5''$. In these models, the recombination radiation is significantly brighter than the I-fronts enclosing the neutral islands.  As a result, it would be difficult to discern neutral island from recombination radiation without a more sophisticated image processing than what is presented here. }
    \label{fig:wide_map_other_models}
\end{figure*}

In Figure \ref{fig:wide_map_other_models} we show the wider-field mock images for the \textsc{$\alpha = 1.5$, slower-$v_\mathrm{IF}$} and \textsc{$\alpha = 1.5$, faster-$v_\mathrm{IF}$} models.  The images are constructed in the same manner as those in Figure \ref{fig:wide_map}, and may be compared to those panels. In contrast to the \textsc{$\alpha = 0.5$, faster-$v_\mathrm{IF}$} model shown in Figure \ref{fig:wide_map}, in these models, the I-fronts are significantly dimmer than the recombination radiation. This characteristic renders the neutral islands more difficult to discern from the recombination radiation smoothed on large scales without prior knowledge of their location.  On the other hand, an image processing approach that is more tailored than the simple smoothing used here could allow the sharp peaks from recombinations to be distinguished from the diffuse large-scale emissions from neutral islands.

\end{document}